1   # Multi-functional foot use during running in the zebra-tailed

2   # lizard (*Callisaurus draconoides*)

3   Chen Li[1], S. Tonia Hsieh[2], and Daniel I. Goldman[1*]

4   [1]*School of Physics, Georgia Institute of Technology, Atlanta, GA 30332, USA and*

5   [2]*Department of Biology, Temple University, Philadelphia, PA 19122, USA*

6   [*]Author for correspondence (daniel.goldman@physics.gatech.edu)

7   **Summary**

8   **A diversity of animals that run on solid, level, flat, non-slip surfaces appear to bounce on**
9   **their legs; elastic elements in the limbs can store and return energy during each step. The**
10  **mechanics and energetics of running in natural terrain, particularly on surfaces that can**
11  **yield and flow under stress, is less understood. The zebra-tailed lizard (*Callisaurus***
12  ***draconoides*), a small desert generalist with a large, elongate, tendinous hind foot, runs**
13  **rapidly across a variety of natural substrates. We use high speed video to obtain detailed**
14  **three-dimensional running kinematics on solid and granular surfaces to reveal how leg, foot,**
15  **and substrate mechanics contribute to its high locomotor performance. Running at ~ 10**
16  **body length/s (~ 1 m/s), the center of mass oscillates like a spring-mass system on both**
17  **substrates, with only 15% reduction in stride length on the granular surface. On the solid**
18  **surface, a strut-spring model of the hind limb reveals that the hind foot saves about 40% of**
19  **the mechanical work needed per step, significant for the lizard's small size. On the granular**
20  **surface, a penetration force model and hypothesized subsurface foot rotation indicates that**
21  **the hind foot paddles through fluidized granular medium, and that the energy lost per step**
22  **during irreversible deformation of the substrate does not differ from the reduction in the**
23  **mechanical energy of the center of mass. The upper hind leg muscles must perform three**
24  **times as much mechanical work on the granular surface as on the solid surface to**
25  **compensate for the greater energy lost within the foot and to the substrate.**

26  Key words: terrestrial locomotion, mechanics, energetics, kinematics, spring-mass system, elastic
27  energy savings, dissipation, granular media

28  Running title: Substrate effects on foot use in lizards

## Introduction

30   Rapid locomotion like running and hopping can be modeled as a spring-mass system bouncing in
31   the sagittal plane (i.e., the Spring-Loaded Inverted Pendulum model, SLIP) (Blickhan, 1989).
32   This has been demonstrated in a variety of animals (Blickhan and Full, 1993; Holmes et al., 2006)
33   in the laboratory on rigid, level, flat, non-slip surfaces (hereafter referred to as "solid surfaces")
34   such as running tracks and treadmills (Dickinson et al., 2000). In the SLIP model, the animal
35   body (represented by the center of mass, CoM) bounces on a single leg (represented by a spring)
36   like a pogo stick, and exerts point contact on the solid ground. The leg spring compresses during
37   the first half of stance, and then recoils during the second half of stance. Through this process, the
38   mechanical (i.e., kinetic plus gravitational potential) energy of the CoM is exchanged with elastic
39   energy stored in the compressed leg spring, reducing energy use during each step. For animals
40   like insects (e.g., Schmitt et al., 2002) and reptiles (e.g., Chen et al., 2006) that run with a
41   sprawled limb posture, the CoM also oscillates substantially in the horizontal plane in a similar
42   fashion, which can also be modeled as a spring-mass system bouncing in the horizontal plane (i.e.,
43   the Lateral Leg Spring model, LLS) (Schmitt et al., 2002). Both the SLIP and the LLS models
44   predict that the mechanical energy of the CoM is lowest at mid-stance and highest during aerial
45   phase.

46   In these models, the spring-mass system and the interaction with the solid ground are perfectly
47   elastic and do not dissipate energy; thus no net work is performed. However, as animals move
48   across natural surfaces, energy is dissipated both within their body and limbs (Fung, 1993) and to
49   the environment (Dickinson et al., 2000). Therefore, mechanisms to reduce energy loss during
50   locomotion can be important. The limbs of many organisms possess elastic elements such as
51   tendons and ligaments that can function as springs to store and return energy during rapid
52   locomotion like running and hopping to decrease energetic cost (Alexander, 2003). Most notable
53   for this function are the ankle extensor tendons in the lower hind leg and the digital flexor
54   tendons and ligaments in the lower fore leg (Alexander, 2003). Furthermore, different limb-
55   ground interaction strategies may be utilized depending on the dissipative properties of the
56   substrate.

57   Laboratory experiments have begun to reveal mechanisms of organisms running on non-solid
58   surbstrates, such as elastic (Ferris et al., 1998; Spence et al., 2010), damped (Moritz and Farley,
59   2003), inclined (Roberts et al., 1997), or uneven (Daley and Biewener, 2006; Sponberg and Full,
60   2008) surfaces; surfaces with few footholds (Spagna et al., 2007); and the surface of water
61   (Glasheen and McMahon 1996a; Hsieh, 2003). While spring-mass-like CoM motion was

63    observed only in some of these studies (Ferris et al., 1998; Moritz and Farley, 2003; Spence et al.,
64    2010), a common finding is that on non-solid surfaces limbs do not necessarily behave like
65    springs to save energy. In addition, these studies suggest that both the active control of body and
66    limb movement through the   nervous system, and the passive mechanical responses of
67    viscoelastic limbs and feet with the environment, play important roles in the control of rapid
68    terrestrial locomotion (for reviews, see Full and Koditschek, 1999; Dickinson et al., 2000).

69    Many substrates found in nature, such as sand, gravel, rubble, dirt, soil, mud, and debris, can
70    yield and flow under stress during locomotion and experience solid-fluid transitions, through
71    which energy may be dissipated via plastic deformation. Understanding locomotion on these
72    substrates is challenging in part because, unlike for flying and swimming where the fluid flows
73    and forces can in principle be determined by solving the Navier-Stokes equations in the presence
74    of moving boundary conditions (Vogel, 1996), no comprehensive force models yet exist for
75    terrestrial substrates that yield and flow (hereafter referred to as "flowing substrates").

76    Granular materials (Nedderman, 1992) like desert sand which are composed of similarly sized
77    particles provide a good model substrate for studying locomotion on flowing substrates.
78    Compared to other flowing substrates, granular materials are relatively simple and the intrusion
79    forces within them can be modeled empirically (Hill et al., 2005). Their mechanical properties
80    can also be precisely and repeatedly controlled using a fluidized bed (Li et al., 2009). In addition,
81    locomotion on granular surfaces is directly relevant for many desert-dwelling reptiles and
82    arthropods such as lizards, snakes, and insects (Mosauer, 1932; Crawford, 1981). Recent
83    advances in the understanding of force and flow laws in granular materials subject to localized
84    intrusion (Hill et al., 2005; Katsuragi and Durian, 2007; Gravish et al., 2010; Ding et al., 2011a)
85    begin to provide insight into the mechanics of locomotion on (and within) granular substrates (Li
86    et al., 2009; Maladen et al., 2009; Mazouchova et al., 2010; Li et al., 2010b; Maladen et al., 2011;
87    Ding et al., 2011b; Li et al., in press).

88    The zebra-tailed lizard (*Callisaurus draconoides*, SVL ~ 10 cm, mass ~ 10 g, Fig. 1A) is an
89    excellent model organism for studying running on natural surfaces, because of its high locomotor
90    performance over diverse terrain. As a desert generalist, this lizard lives in a range of desert
91    habitats including flat land, washes, and sand dunes (Vitt and Ohmart, 1977; Korff and McHenry,
92    2011), and encounters a large variety of substrates ranging from rocks, gravel, closely-packed
93    coarse sand, and loosely-packed fine sand (Karasov and Anderson, 1998; Korff and McHenry,
94    2011). The zebra-tailed lizard is the fastest-running species among desert lizards of similar size

95    (Irschick and Jayne, 1999a), and has been observed to run at up to 4 m/s (50 bl/s) both on solid
96    (e.g., treadmill) (Irschick and Jayne, 1999a) and on granular (e.g., sand dunes) (Irschick and
97    Jayne, 1999b) surfaces. Its maximal acceleration and running speed also did not differ
98    significantly when substrate changes from coarse wash sand to fine dune sand, whose yield
99    strengths differ by a factor of three (Korff and McHenry, 2011).

100   Of particular interest is whether and how the zebra-tailed lizard's large, elongate hind foot
101   contributes to its high locomotor capacity. In addition to a slim body, a long tapering tail, and
102   slender legs (Fig. 1A), the zebra-tailed lizard has an extremely large, elongate hind foot, the
103   largest (40% SVL) among lizards of similar size (Irschick and Jayne, 1999a). Its hind foot is
104   substantially larger than the fore foot (area = 1 cm$^2$ vs. = 0.3 cm$^2$) and likely plays a dominant role
105   for locomotion (Mosauer, 1932). Recent studies in insects, spiders, and geckos (Jindrich and Full,
106   1999; Antumn et al., 2000; Dudek and Full, 2006; Spagna et al., 2007) suggested that animals can
107   rely on appropriate morphology and material properties of their bodies and limbs to accommodate
108   variable, uncertain conditions during locomotion. Despite suggestions that the large foot area
109   (Mosauer, 1932) and increased stride length via elongate toes may confer locomotor advantages
110   (Irschick and Jayne, 1999a), the mechanisms of how the hind foot contributes to the zebra-tailed
111   lizard's high running capacity remain unknown.

112   In this paper, we study the mechanics and mechanical energetics of the zebra-tailed lizard running
113   on two well-defined model surfaces, a solid surface and a granular surface. These two surfaces lie
114   on opposite ends of the spectrum of substrates that the zebra-tailed lizard encounters in its natural
115   environment, and present distinct conditions for locomotion. We investigate whether the lizard's
116   center of mass bounces like a spring-mass system during running on both solid and granular
117   surfaces. We combine measurements of three-dimensional kinematics of the lizard's body, hind
118   limb, and hind foot, dissection and resilience measurements of the hind limb, and modeling of
119   foot-ground interactions on both substrates, and demonstrate that the lizard's large, elongate hind
120   foot serves different functions during running on solid and granular surfaces. We find that on the
121   solid surface, the hind foot functions as an energy-saving spring; on the granular surface, it
122   functions as a dissipative, force-generating paddle to generate sufficient lift during each step. The
123   larger energy dissipation to the substrate and within the foot during running on the granular
124   surface must be compensated for by greater mechanical work done by the upper hind leg muscles.

125

126                                           **Materials and methods**

127                                          *Animals*

128    Seven adult zebra-tailed lizards (*Callisaurus draconoides*) were collected from the Mojave Desert,

129    AZ, USA in 08/2007 (Permit SP591773) for three-dimensional kinematics experiments. Table 1

130    shows the morphological measurements for these seven animals. Eleven additional adult animals

131    were collected from the Mojave Desert, CA, USA in 09/2009 (Permit SC 10901) for hind limb

132    resilience measurements. Two preserved specimens were used for dissection. The animals were

133    housed in the Physiological Research Laboratory animal facility of The Georgia Institute of

134    Technology. Each animal was housed individually in an aquarium filled with sand, and fed

135    crickets and mealworms dusted with vitamin and calcium supplement two to three times a week.

136    The ambient temperature was maintained at 28°C during the day and 24°C during the night. Full-

137    spectrum fluorescent bulbs high in UVB were set to a 12 hour/12 hour light/dark schedule.

138    Ceramic heating elements were provided 24 hours a day to allow the animals to thermo-regulate

139    at preferred body temperature. All experimental procedures were conducted in accordance with

140    The Georgia Institute of Technology IACUC protocols.

141                                      *Surface treatments*

142    A wood board ($120 \times 23 \times 1$ cm$^3$) bonded with sandpaper (grit size $\sim$ 0.1 mm) for enhanced

143    traction was used as the solid surface. Glass particles (diameter = $0.27 \pm 0.04$ mm mean $\pm$ 1

144    standard deviation, density = $2.5 \times 10^3$ kg/m$^3$, Jaygo Incorporated, Union, NJ, USA) were used as

145    the granular substrate, which are approximately spherical and of similar size to typical desert sand

146    (Dickinson and Ward, 1994). Before each trial, a custom-made fluidized bed trackway (200 cm

147    long, 50 cm wide) prepared the granular substrate (12 cm deep) into a loosely packed state

148    (volume fraction = 0.58) for repeatable yield strength (for experimental details of the fluidized

149    bed trackway, see Li et al., 2009).

150                                  *Three-dimensional kinematics*

151    We used high speed video to obtain three-dimensional kinematics as the lizard ran across the

152    prepared surfaces (Fig. 1B). Before each session, high-contrast markers (Wite-Out, Garden Grove,

153    CA, USA) were painted on each animal for digitizing at nine joints along the midline of the trunk

154    and the right hind limb (Fig. 1A,B): neck (N), center of mass (CoM), pelvis (P), hip (H), knee (K),

155    ankle (A); and the metatarsal-phalangeal joint (MP), distal end of the proximal phalanx (PP), and

156    digit tip (T) of the fourth toe. The approximate longitudinal location of the CoM in resting

157    position was determined by tying a thread around the body of an anesthetized lizard and

158 repositioning the thread until the body balanced horizontally. Before each trial, the surface was
159 prepared (for the granular surface treatment only), and calibration images were taken of a custom-
160 made 39-point calibration object (composed of LEGO, Billund, Denmark). The animal was then
161 induced to run across the field of view by a slight tap or pinch on the tail. Two synchronized AOS
162 high speed cameras (AOS Technologies, Baden Daettwil, Switzerland) captured simultaneous
163 dorsal and lateral views at 500 frame/s (shutter time = 300 μs). The ambient temperature was
164 maintained at 35°C during the test. Animals were allowed to rest at least five minutes between
165 trials and at least two days between sessions.

166 We digitized the calibration images and high speed videos, and used direct linear transformation
167 (DLT) to reconstruct three-dimensional kinematics from the two-dimensional kinematics from
168 both dorsal and lateral views. Digitization and DLT calculations were performed using custom
169 software (DLTcal5 and DLTdv5, Hedrick, 2008). Axes were set such that $+x$ pointed in the
170 direction of forward motion, $+z$ pointed vertically upward, and $+y$ pointed to the left of the animal.
171 Footfall patterns of touchdown and takeoff were determined from the videos. On the granular
172 surface, because the hind foot often remained obscured by splashed grains during foot extraction,
173 we defined foot takeoff as when the knee began to flex following extension during limb
174 protraction (which is when foot takeoff occurs on the solid surface). To reduce noise and enable
175 direct comparisons among different running trials, position data were filtered with a Butterworth
176 low-pass filter with a cutoff frequency of 75 Hz, and interpolated to 0–100% of one full stride
177 period ($T$) between two successive touchdowns of the right hind limb. All data analysis was
178 completed with MATLAB (MathWorks, Natick, MA, USA) unless otherwise specified.

179 *Statistics*

180 We accepted trials that met the following criteria: the animal ran continuously through the field of
181 view, the run was straight without contacting sidewalls of the trackway, there was a full stride
182 (between two consecutive touchdowns of the right hind limb) in the range of view, all the nine
183 markers were visible throughout the full stride, and the forward speed changed less than 20%
184 after the full stride. With these criteria, out of a total of 125 trials from 7 individuals on both solid
185 (61 trials from 7 individuals) and granular (64 trials from 7 individuals) surfaces collected over a
186 period of over three months, we ultimately accepted 51 runs from 7 individuals on solid (23 runs
187 from 7 individuals) and granular (28 runs from 7 individuals) surfaces. Because the data set had
188 an unequal number of runs per individuals, and because we were measuring freely-running
189 animals and did not control for speed, to maintain statistical power, all statistical tests were

190  performed on a subset of these data using one representative run per individual on both solid ($N =$
191  7) and granular ($N = 7$) surfaces. The representative run for each individual was selected based on
192  having the most consistent running speed for at least one full stride and was also closest to the
193  mean running speed of all 51 trials. Data are reported as mean ± 1 standard deviation (s.d.) from
194  the 7 representative runs on each substrate unless otherwise specified.

195  To determine the effect of substrate, all kinematic variables were corrected for size-related
196  differences by regressing the variables against SVL and taking the residuals for those that
197  regressed significantly with SVL ($P < 0.05$). We then ran an ANCOVA with substrate and speed
198  as covariates to test for substrate effects, independent of running speeds. All statistical tests were
199  performed using JMP (SAS, Cary, NC, USA).

200  For the energetics data, we used dimensionless quantities by normalizing energies of each run to
201  the CoM mechanical energy at touchdown of that run, thus eliminating the effect of mass and
202  running speed on energies. An ANOVA was used to test the differences between the reduction in
203  CoM mechanical energy, elastic energies, and energy loss. A Tukey's HSD was used for post-hoc
204  tests where needed.

*Dissection and model of hind limb*

206  To gain insight into the role of anatomical components of the hind limb on mechanics during
207  locomotion, we dissected the hind limb of two preserved specimens. We quantified anatomical
208  dimensions by measuring the radii of the knee (K), ankle (A), the metatarsal-phalangeal joint
209  (MP), the distal end of the proximal phalanx (PP), and the digit tip (T) of the fourth toe. We also
210  observed the muscle and tendon arrangements within the lower leg and the foot. Based on these
211  anatomical features, we developed a model of the hind limb which incorporated the structure,
212  properties, and function of its main elements.

*Resilience measurements of hind limb*

214  To characterize the resilience of the hind limb for estimation of energy return, a modification of
215  the work loop technique was used (Fig. 2A), in which the limb was kept intact and forces were
216  applied to the whole limb instead of a single muscle (Dudek and Full, 2006). The animal was
217  anesthetized using 2% isoflurane air solution during the test. The hind foot was maintained within
218  the vertical plane, pushed down onto and then extracted from a custom force platform suited for
219  small animals (10.2 × 7.6 cm², range = 2.5 N, resolution = 0.005 N) bonded with sandpaper (grit
220  size ~ 0.1 mm). Ground reaction force $F$ was measured at 10 kHz sampling rate using a custom

221  LabVIEW program (National Instruments, Austin, TX, USA). A Phantom high speed camera

222  (Vision Research, Wayne, NJ, USA) simultaneously recorded deformation of the foot from the

223  side view at 250 frame/s (shutter time = 500 μs). High-contrast markers (Wite-Out, Garden Grove,

224  CA, USA) were painted on the joints of the hind foot (A, MP, PP, T, and a point on the tibia

225  above the ankle). The ambient temperature was maintained at 35°C during the test.

226  Videos of foot deformation were digitized to obtain the angular displacement of the foot $\Delta\theta$ =

227  $\theta_0 - \theta_f$, i.e., the change in the angle formed by the tibia and the foot (from the ankle to the digit tip

228  of the fourth toe) (Fig. 2A). Angular displacement $\Delta\theta$ was synchronized with the measured torque

229  $\tau$ about the ankle (calculated from the measured ground reaction force) to obtain a passive work

230  loop. The damping ratio of the hind limb, i.e., the percentage of energy lost within the hind limb

231  after loading and unloading, was calculated as the fraction of area within a work loop relative to

232  the area under the higher loading curve (Fung, 1993). Hind limb resilience, i.e., the percentage of

233  energy returned by the foot after loading and unloading, was one minus the damping ratio (Ker et

234  al., 1987; Dudek and Full, 2006). An ANOVA was used to test the effect of maximal torque,

235  maximal angular displacement, loading rate, and individual animal on hind limb resilience.

236  *Granular penetration force measurements*

237  While comprehensive force models are still lacking to calculate ground reaction forces during

238  locomotion granular media, a low speed penetration force model was previously used to explain

239  the locomotor performance of a legged robot on granular media (Li et al., 2009). Similarly, to

240  estimate the vertical ground reaction force on the lizard foot during running on the granular

241  surface, we measured the vertical force on a plate slowly penetrating vertically into the granular

242  substrate (Fig. 2B). Before each trial, a fluidized bed (area = $24 \times 22$ cm$^2$) prepared the granular

243  substrate (depth = 12 cm) into a loosely packed state (volume fraction = 0.58) (for details, see

244  Maladen et al., 2009). A robotic arm (CRS robotics, Burlington, OT, Canada) pushed a

245  horizontally-oriented plate vertically downward at 0.01 m/s into the granular substrate to a depth

246  of 7.6 cm, and then extracted the plate vertically at 0.01 m/s. The force on the plate was measured

247  by a force transducer (ATI Industrial Automation, Apex, NC, USA) mounted between the robotic

248  arm and the plate at 100 Hz sampling rate using a custom LabVIEW program (National

249  Instruments, Austin, TX, USA). The depth of the plate was measured by tracking the position of

250  an LED light marker mounted on the robotic arm in side view videos taken by a Pike high speed

251  camera (Edmund Optics, Barrington, NJ, USA). Two thin aluminum plates of different area were 252    used ($A_1 = 7.6 \times 2.5$ cm$^2$ and $A_2 = 3.8 \times 2.5$ cm$^2$; thickness = 0.6 cm). Three trials were performed

253    for each plate.

254

255                      **Results**

256                     *Performance and gait*

257    On both solid and granular surfaces, the zebra-tailed lizard ran with a diagonal gait, a sprawled

258    limb posture, and lateral trunk bending (see Fig. 3 and Movies 1, 2 in supplementary material for

259    representative runs on both substrates). Figure 4 shows average forward speed $\bar{v}_{x,\text{CoM}}$, stride

260    frequency $f$, and duty factor $D$ of the entire data set (all symbols; 23 runs on the solid surface and

261    28 runs on the granular surface) and of the representative runs (filled symbols; $N = 7$ on the solid

262    surface and $N = 7$ on the granular surface). Table 2 lists mean values and statistical results for all

263    the gait and kinematic variables from the representative runs for both solid ($N = 7$) and granular

264    ($N = 7$) surfaces. On both surfaces, $\bar{v}_{x,\text{CoM}}$ increased with $f$ (Fig. 4A, $P < 0.05$, ANCOVA), and $D$

265    decreased with $\bar{v}_{x,\text{CoM}}$ (Fig. 4B, $P < 0.05$, ANCOVA). $D \approx 0.45$ on both surfaces resulting in an

266    aerial phase of approximately 5% stride period $T$ between alternating stances (Fig. 5A). Neither

267    $\bar{v}_{x,\text{CoM}}$ ($P > 0.05$, ANOVA) nor $D$ ($P > 0.05$, ANCOVA) significantly differ between surfaces.

268    Average stride length $\lambda = \bar{v}_{x,\text{CoM}}/f$ was 15% shorter on the granular surface ($P < 0.05$, ANCOVA).

269                  *Center of mass kinematics*

270    The lizard displayed qualitatively similar center of mass oscillations during running on both

271    surfaces. The CoM forward speed $v_{x,\text{CoM}}$ (Fig. 5B) and vertical position $z_{\text{CoM}}$ (Fig. 5C) oscillated at

272    $2f$, dropping during the first half and rising during the second half of a stance, i.e., reaching

273    minimum at mid-stance and maximum during the aerial phase. The CoM also oscillated medio-

274    laterally at $f$ (Fig. 5D). Throughout the entire stride, $z_{\text{CoM}}$ was significantly higher on the solid

275    surface ($P < 0.05$, ANCOVA). The CoM vertical oscillations $\Delta z_{\text{CoM}}$ and lateral oscillations $\Delta y_{\text{CoM}}$

276    did not differ between substrates ($P > 0.05$, ANCOVA).

277              *Hind foot, hind leg, and trunk kinematics*

278    The lizard displayed distinctly different hind foot, hind leg, and trunk kinematics during running

279    on solid and granular surfaces (Figs. 3, 6). On the solid surface, the lizard used a digitigrade foot

280    posture (Fig. 3A–E, solid line/curve). During the entire stride, the hind foot engaged the solid

281   surface only with the digit tips. At touchdown, the toes were straight and pointed slightly
282   downward. The touchdown foot angle (measured along the fourth toe) was $\theta_{touchdown} = 12 \pm 4°$
283   relative to the surface (Fig. 3A,E; Fig. 6A, red). During stance, the long toes pivoted over the
284   stationary digit tips (Fig. 3A−C, vertical dotted line shows zero displacement) and hyperextended
285   into a c-shape (Fig. 3B, solid curve). The foot straightened again at takeoff, pointing downward
286   and slightly backward (Fig. 3C, solid line), and then flexed during swing (Fig. 3D, solid curve).

287   On the granular surface, the lizard used a plantigrade foot posture (Fig. 3F,J, solid line). At
288   touchdown, the hind foot was nearly parallel with the surface, with the toes spread out and held
289   straight. In the vertical direction, the foot impacted the granular surface at speeds of up to 1 m/s.
290   The ankle joint slowed down to ∼ 0.1 m/s within a few milliseconds following impact (a few
291   percent of stride period $T$) while the the foot started penetrating the surface. The touchdown foot
292   angle was $\theta_{touchdown} = 4 \pm 3°$ relative to the surface (Fig. 3J; Fig. 6A, blue), significantly smaller
293   than that on the solid surface ($P < 0.05$, ANCOVA). During stance, the entire foot moved
294   subsurface and was obscured (Fig. 3G). The ankle joint remained visible right above the surface
295   and moved forward by about a foot length (Fig. 3F−H, vertical dotted line shows ankle
296   displacement). The foot was extracted from the substrate at takeoff, pointing downward and
297   slightly backward, and then flexed during swing (Fig. 3I, solid curve).

298   As a result of foot penetration on the granular surface, both the knee height $z_{knee}$ (Fig. 6B) and
299   pelvis height $z_{pelvis}$ (Fig. 6C) were lower on the granular surface ($P < 0.05$, ANCOVA). In addition,
300   on the granular surface, the knee moved downward by a larger vertical displacement $\Delta z_{knee}$ during
301   the first half of stance ($P < 0.05$, ANCOVA; Fig. 6B), while the knee joint extended by a larger
302   angle $\Delta\theta_{knee}$ during the second half of stance ($P < 0.05$, ANCOVA; Fig. 6D). Throughout the
303   entire stride, the trunk was nearly horizontal on the solid surface (Fig. 3A−D, dashed line), but
304   pitched head-up on the granular surface (Fig. 3F−I, dashed line; Fig. 6E). On both surfaces, the
305   hind legs were sprawled at an angle of $\theta_{sprawl} \approx 40°$ during stance (Fig. 3; $\theta_{sprawl}$ is defined as the
306   angle between the horizontal plane and the leg orientation in the posterior view). In most runs, the
307   tail was farther from the solid surface and closer to the granular surface (Fig. 3).

*Hind limb anatomy*

309   From morphological measurements (Table 1), the hind foot of the zebra-tailed lizard comprised
310   42% of the hind limb length, and the longest fourth toe alone accounted for 63% of the hind foot
311   length. These ratios are in similar range to previous observations (Irschick and Jayne, 1999a). The

312    slender foot had a cross-sectional radius of $r = 0.50-1.25$ mm tapering distally, with reducing
313    joint radii: $r_K = r_A = 1.25$ mm, $r_{MP} = 0.75$ mm, $r_{PP} = r_T = 0.50$ mm.

314    Unlike many cursorial mammals whose ankle extensor muscles of the lower hind leg have long
315    tendons (Alexander, 2003), ankle extensor tendons are nearly non-existent in the zebra-tailed
316    lizard (Fig. 7A). Instead, layers of elongate tendons were found in both the dorsal and ventral
317    surfaces of the foot. Our anatomical description is focused on the ventral muscle and tendon
318    anatomy in the hind limb and terms given to muscles and tendons follow (Russell, 1993). A large,
319    tendinous sheath, the superficial femoral aponeurosis, originates from the femoro-tibial
320    gastrocnemius, stretches across the ventral surface of the foot, and inserts on the metatarsal-
321    phalangeal joints for digits III and IV. The superficial portion of the femoro-tibial gastrocnemius
322    muscle body extends to the base of the ankle, thereby rendering the human equivalent of the
323    ankle extensor tendons (i.e., the "Achilles" tendon) absent. Deep to the superficial femoral
324    aponeurosis lie the flexor digitorum brevis muscles (not shown) which control the flexion of each
325    of the digits. Tendons from the flexor digitorum longus muscle located on the lower hind leg run
326    deep to the flexor digitorum brevis muscle bodies, and extend to the tips of the digits. No
327    additional tendons are visible deep to the flexor digitorum longus tendons.

328    *Hind limb model*

329    Based on the observed muscle and tendon anatomy, we propose a two-dimensional strut-spring
330    model of the hind limb (Fig. 7B), which assumes isometric contraction for the lower leg muscles
331    and incorporates the spring nature of the foot tendons. This model is inspired from previous
332    observations in large running and hopping animals of the strut-like function of ankle extensor
333    muscles (Biewener, 1998a; Roberts et al., 1997) and spring-like function of ankle extensor
334    tendons (for a review, see Alexander, 2003). Rigid segments (Fig. 7B, dashed lines), which are
335    free to rotate about joints within a plane, represent the skeleton. The ankle extensor muscles in the
336    lower leg, which originate on the femur and run along the ventral side of the tibia, are modeled as
337    a rigid strut (muscle strut, Fig. 7B, blue line) that contracts isometrically during stance in running.
338    A linear spring (tendon spring, Fig. 7B, red line), which originates from the distal end of the
339    muscle strut and extends to the digit tip, models the elastic foot tendons. The muscle strut and
340    tendon spring are ventrally offset from the midline of the skeleton at each joint by respective joint
341    radii.

342    *Hind limb resilience*

343　Representative passive work loops (Fig. 8A–C) showed that torque τ was higher when the foot

344　was pushed down on the solid surface than when it was extracted, similar to  previous

345　observations in humans (Ker et al., 1987) and cockroaches (Dudek and Full, 2006). Maximal

346　torque was positively correlated with maximal angular displacement ($F_{1,62}$ = 64.3188, $P$ < 0.001,

347　ANOVA). The kinks observed in the middle of the loading curve were due to the fifth toe

348　contacting the surface. Average hind limb resilience calculated from the work loops was $R$ = 0.44

349　± 0.12 (Fig. 8D–F, 3 individuals, 64 trials). $R$ did not differ between individuals ($F_{2,61}$ = 2.1025, $P$

350　= 0.1309, ANOVA), and did not depend on maximal torque ($F_{1,62}$ = 0.5208, $P$ = 0.4732, ANOVA;

351　Fig. 8D), maximal angular displacement ($F_{1,62}$ = 0.0164, $P$ = 0.8987, ANOVA; Fig. 8E), or

352　average loading rate ($F_{1,62}$ = 1.1228, $P$ = 0.2934, ANOVA; Fig. 8F).

353　　　　　　　　　*Hind foot curvature, tendon deformation, and tendon stiffness*

354　The observed three-dimensional positions of the hind limb fit well to the two-dimensional hind

355　limb model (Fig. 9A–D), and enabled calculation of the curvature, tendon deformation, and

356　tendon stiffness of the hind foot (see Appendix). Calculated hind foot curvature κ (Fig. 9E, solid

357　curve) showed that the hind foot hyperextended during stance (positive κ) and flexed during

358　swing (negative κ). The foot was straight at touchdown and shortly after takeoff (κ = 0).

359　Calculated tendon spring deformation Δ*l* (Fig. 9E, dashed curve) showed that the tendon spring

360　stretched during the first half and recoiled during the second half of stance. The estimated tendon

361　spring stiffness was $k = 4.4 \times 10^3$ N/m (see Appendix).

362　　　　　　　　　　　　*Mechanical energetics on solid surface*

363　Using the observed CoM and hind limb kinematics, calculated tendon spring stiffness and

364　deformation, and measured hind limb resilience, we examined the mechanical energetics of the

365　lizard running on the solid surface (Table 3, Fig. 9F). From the observed CoM kinematics, in the

366　first half of stance, the mechanical energy of the CoM (kinetic energy plus gravitational potential

367　energy) decreased significantly from $E_{\text{touchdown}}$ = 1.00 ± 0.00 at touchdown to $E_{\text{mid-stance}}$ = 0.81 ±

368　0.08 at mid-stance ($F_{2,18}$ = 12.2345, $P$ = 0.0004, ANOVA, Tukey HSD). In the second half of

369　stance, the mechanical energy of the CoM recovered to $E_{\text{aerial}}$ = 0.95 ± 0.10 at mid aerial phase,

370　not significantly different from $E_{\text{touchdown}}$ (Tukey HSD). The reduction in CoM mechanical energy

371　in the first half of stance $\Delta E_{\text{mech}}$ = 0.19 ± 0.08 is the mechanical work needed per step on the solid

372　surface. Note that the energies of each run were normalized to $E_{\text{touchdown}}$ of that run.

373    At mid-stance, the elastic energy stored in the tendon spring was $E_{storage} = 0.18 \pm 0.13$ (calculated

374    from $1/2\ k\Delta l_{max}^{2}$, see Appendix), not significantly different from $\Delta E_{mech}$ ($F_{1,12} = 0.0475$, $P =$

375    $0.8312$, ANOVA). Because hind limb resilience $R = 0.44 \pm 0.12$, the elastic recoil of the foot

376    tendons returned an energy of $E_{return} = RE_{storage} = 0.08 \pm 0.06$, or $41 \pm 33\%$ of the mechanical work

377    needed per step ($\Delta E_{mech}$) on the solid surface. We verified that foot flexion during swing induced

378    little energy storage ($< 0.1\ E_{storage}$) because the hind foot was less stiff during flexion ($0.7 \times 10^{3}$

379    N/m) than during hyperextension ($4.4 \times 10^{3}$ N/m).

380                                         *Granular penetration force model*

381    Although little is known about the kinematics and mechanics of the complex limb intrusions

382    during legged locomotion on granular surfaces, we took inspiration from previous observations

383    that horizontal drag (Maladen et al., 2009) and vertical impact (Katsuragi and Durian, 2007)

384    forces in glass particles were insensitive to speed when intrusion speed was below approximately

385    0.5 m/s. Because the kinematics observed on the granular surface suggest that the vertical speeds

386    of most of the foot relative to the ground were below 0.5 m/s during most of the stance phase (see

387    Appendix), we assumed that the ground reaction forces on the lizard's feet were also insensitive

388    to speed. This allowed us to use the vertical penetration force measured at 0.01 m/s to model and

389    estimate the vertical ground reaction forces on the lizard foot.

390    From the force data on both plates (Fig. 10), vertical ground reaction force $F_z$ was proportional to

391    both penetration depth $|z|$ and projected area $A$ of the plate (area projected into the horizontal

392    plane). $F_z$ was pointing upward during foot penetration, and pointing downward during foot

393    extraction and dropped by an order of magnitude. These measurements were in accord with

394    previous observations of forces on a sphere penetrating into granular media (Hill et al., 2005).

395    Furthermore, we estimated from free falling of particles under gravity that it would take longer

396    than the stance duration (45 ms) for the grains surrounding a penetrating foot to refill a hole

397    created by the foot of maximal depth ($|z|_{max} = 1.0$ cm, see Appendix). Thus we assumed that the

398    vertical ground reaction forces were negligible during foot extraction.

399    Therefore, we approximate the vertical penetration force as:

400    
$$F_z = \begin{cases} \alpha|z|A, & \text{for increasing } |z|, \\ 0, & \text{for decreasing } |z|, \end{cases} \qquad (1)$$

401    where $\alpha$ is the vertical stress per unit depth, which is determined by the properties of the granular

402    material and increases with compaction (Li et al., 2009). Fitting $F_z = \alpha|z|A$ to the force data

403    during penetration over regions where the plate was fully submerged and far from boundary (Fig.

404    10, dashed lines), we obtained $\alpha = 3.5 \times 10^5$ N/m$^3$ for loosely packed $0.27 \pm 0.04$ mm diameter

405    glass particles.

406    *Vertical ground reaction force on granular surface*

407    During a stance on the granular surface, the CoM vertical speed $v_{z,\mathrm{CoM}}$ (calculated from $z_{\mathrm{CoM}}$) was

408    approximately sinusoidal (Fig. 11A, dashed curve). This implies that the $F_z$ on a lizard foot must

409    be approximately sinusoidal. In addition, the foot was nearly horizontal at touchdown, but pointed

410    downward and slightly backward during takeoff. In consideration of the functional form of the

411    penetration force (Eqn. 1), we hypothesized that during stance the foot rotated subsurface by $\pi/2$

412    in the sagittal plane (Fig. 11C), increasing foot depth $|z|$ but decreasing projected foot area $A$, thus

413    resulting in a sinusoidal $F_z$ which reaches a maximum at mid-stance before the foot reaches

414    largest depth (see Appendix). A sinusoidal $F_z$ is also possible for a fixed projected foot area if the

415    foot maintains contact on solidified grains. However, this is unlikely considering that during

416    stance the ankle moved forward at the surface level by a foot length.

417    Assuming that during stance the hind foot rotated by $\pi/2$ in the sagittal plane at a constant angular

418    velocity, the vertical ground reaction force that each foot generated was $F_z = 5\pi mg/9 \, \sin 10\pi t/9T$

419    (see Appendix). The net vertical acceleration due to this $F_z$ and the animal weight $mg$ was $a_z =$

420    $F_z/\mathrm{m} - \mathrm{g}$ (Fig. 11B; solid and dashed curves are $a_z$ from both hind feet, shifted from each other by

421    $T/2$). The CoM vertical speed $v_{z,\mathrm{CoM}}$ predicted from the total $a_z$ on both hind feet (Fig. 11A,

422    dashed curve) agreed with experimental observations (Fig. 11A, solid curve). The slight under-

423    prediction of the oscillation magnitudes of $v_{z,\mathrm{CoM}}$ was likely due to an over-estimation of duty

424    factor on the granular surface. This is because $F_z$ may have dropped to zero even before takeoff if

425    the foot started moving upward before takeoff (Fig. 10).

426    *Mechanical energetics on granular surface*

427    Using the measured CoM kinematics, assumed foot rotation, and calculated vertical ground

428    reaction force, we examined the mechanical energetics of the lizard running on the granular

429    surface (Table 3, Fig. 11D). In the first half of stance, the mechanical energy of the CoM

430    decreased significantly from $E_{\mathrm{touchdown}} = 1.00 \pm 0.00$ at touchdown to $E_{\mathrm{mid\text{-}stance}} = 0.86 \pm 0.09$ at

431    mid-stance ($F_{2,18} = 6.6132$, $P = 0.007$, ANOVA, Tukey HSD). In the second half of stance, the

432    mechanical energy of the CoM recovered to $E_{\mathrm{aerial}} = 0.99 \pm 0.10$ at mid aerial phase, not

433    significantly different from $E_{\mathrm{touchdown}}$ (Tukey HSD). The reduction in CoM mechanical energy in 434    the first half of stance $\Delta E_{mech} = 0.14 \pm 0.09$ is the mechanical work needed per step on the

435    granular surface. By integration of $F_z$ over vertical displacement of the foot during stance (see

436    Appendix), the energy lost to the granular substrate per step was estimated as $E_{substrate} = 0.17 \pm$

437    0.05, not significantly different from $\Delta E_{mech}$ ($F_{1,12} = 0.4659$, $P = 0.5078$, ANOVA). Note that the

438    energies of each run were normalized to $E_{touchdown}$ of that run.

439

440                         **Discussion**

441          *Conservation of spring-mass-like CoM dynamics on solid and granular surfaces*

442    The observed kinematics and calculated mechanical energetics demonstrated that the zebra-tailed

443    lizard ran like a spring-mass system on both solid and granular surfaces. On both surfaces, the

444    CoM forward speed (Fig. 5B), vertical position (Fig. 5C), and lateral position (Fig. 5D) displayed

445    oscillation patterns that are in accord with predictions from the Spring-Loaded Inverted Pendulum

446    (SLIP) model (Blickhan, 1989) and the Lateral Leg Spring (LLS) model (Schmitt et al., 2002).

447    The small relative oscillations of the CoM forward speed (i.e., $\Delta v_{x,CoM}/v_{x,CoM} \ll 1$) was expected

448    because the Froude number was large for the lizard (see Appendix). The substantial sprawling of

449    the legs contributed to the medio-lateral oscillatory motion of the animal. Furthermore, on both

450    surfaces, the mechanical energy of the CoM oscillated within a step, reaching minimum at mid-

451    stance and maximum during the aerial phase (Fig. 9F, 11D), a defining feature of spring-mass

452    like running (Blickhan, 1989; Schmitt et al., 2002).

453    To our knowledge, ours is the first study to quantitatively demonstrate spring-mass-like CoM

454    motion in lizards running on granular surfaces. Spring-mass-like CoM motion was previously

455    observed in other lizards and geckos running on solid surfaces (Farley and Ko, 1997; Chen et al.,

456    2006), but it was not clear whether energy-saving by elastic elements played an important role.

457          *Hind foot function on solid surface: energy-saving spring*

458    Our study is also the first to quantify elastic energy savings in foot tendons in lizards during

459    running on solid surfaces. The significant energy savings (about 40% of the mechanical work

460    needed per step) in the zebra-tailed lizard's hind foot tendons is in a similar range to the energy

461    savings by ankle extensor tendons and digital flexor tendons and ligaments in larger animals

462    (Alexander, 2003), such as kangaroos (50%, Alexander and Vernon, 1975), wallabies (45%,

463 Biewener et al., 1998), horses (40%, Biewener, 1998b), and humans (35%, with an additional 17%
464 from ligaments in the foot arch, Ker et al., 1987).

465 This is surprising considering that the elastic energy saving mechanism was previously thought
466 less important in small animals (e.g., 14% in hopping kangaroo rat of ∼ 100 g mass, Biewener et
467 al., 1981). Because the tendons of small animals are "overbuilt" to withstand large stresses during
468 escape, during steady-speed locomotion these tendons usually experience stresses too small to
469 induce significant elastic energy storage and return (Biewener and Blickhan, 1988; McGowan et
470 al., 2008). We verified that for zebra-tailed lizards running at ∼ 1 m/s the maximal stress in the
471 foot tendons is 4.3 MPa (see Appendix), well below the 100 MPa breaking stress for most
472 tendons (Kirkendall and Garrett, 1997).

473 The zebra-tailed lizard's elongate hind foot and digitigrade foot posture on the solid surface may
474 be an adaptation for elastic energy saving during rapid locomotion. Like other iguanids (Russell,
475 1993), this lizard does not have substantial ankle extensor tendons as large animals do.
476 Nevertheless, elongation of foot tendons and a digitigrade posture enhance the hind foot's energy
477 saving capacity by decreasing tendon stiffness and mechanical advantage (Biewener et al., 2004)
478 (see Appendix). A recent study also found significant energy savings (53%) by elongate foot
479 tendons in running ostriches (Rubenson et al., 2011). More generally, elongation of distal limb
480 segments such as legs, feet, and toes which possess tendons may be an adaptation for energy
481 saving during rapid locomotion. Indeed, many cursorial animals including mammals (Garland Jr.
482 and Janis, 1993), lizards (Bauwens et al., 1995), and dinosaurs (Coombs Jr., 1978) display
483 elongation of distal limb segments. Short fascicles and long tendons and ligaments are often
484 found in the ankle extensor muscles and digital flexor muscles in large cursorial ungulates such as
485 horses, camels, and antelopes (Alexander, 2003).

*Solid surface model assumptions*

487 Our estimates of elastic energy storage and return on the solid surface assume isometric
488 contraction of lower leg muscles. However, muscles have a finite stiffness and do lengthen by a
489 small amount under limb tension (Biewener, 1998a; Roberts et al., 1997). Despite this difference,
490 our estimates still hold, because in the latter case both lower leg muscles and foot tendons behave
491 like springs, and the total stiffness remains the same (since external force and total deformation
492 remain the same). In the case where the muscles actively shorten during stance and further
493 lengthen the tendons (which does positive mechanical work on the tendons), the energy storage
494 and return in the tendons would increase. However, the overall energy efficiency would decrease

495 (with everything else being the same), because apart from energy lost in tendon recoil, energy is
496 further lost in the muscles that perform the mechanical work, i.e., muscle work is more expensive
497 than tendon work (Biewener and Roberts, 2000).

498 In addition, the hind limb resilience obtained from anesthetized lizards was assumed to be a good
499 estimate for hind limb resilience in running lizards. This is based on our observations that
500 resilience was independent of torque, angular displacement, and loading rate, as well as previous
501 findings that the damping properties of animal limbs are largely intrinsic to their structure and
502 material properties (Weiss et al., 1988; Fung, 1993; Dudek and Full, 2006). Future studies using
503 techniques such as tendon buckles (Biewener et al., 1998), sonomicrometry (Biewener et al.,
504 1998), ultrasonography (Maganaris and Paul, 1999), and oxygen consumption measurement
505 (Alexander, 2003) during locomotion are needed to confirm this assumption.

506 *Hind foot function on granular surface: dissipative, force-generating paddle*

507 The similarity between the observed and predicted $v_{z,CoM}$ on the granular surface supports the
508 hypothesis of subsurface foot rotation. We speculate that on the granular surface the foot
509 functions as a "paddle" through fluidized grains to generate force. This differs from previous
510 observations of the utilization of solidification forces of the granular media in a legged robot (Li
511 et al., 2009; Li et al., 2010b) and sea turtle hatchlings (Mazouchova et al., 2010) moving on
512 granular surfaces.

513 As the zebra-tailed lizard's hind foot paddles through fluidized grains to generate force, energy is
514 lost to the substrate because grain contact forces in granular media are dissipative (Nedderman,
515 1992). A large foot can reduce energy loss to the granular substrate compared to a small one,
516 much like large snowshoes used by humans can reduce energy cost for walking on snow (Knapik
517 et al., 1997). From our model of foot-ground interaction on the granular surface, for a given
518 animal (constant weight), energy loss to the substrate is proportional to foot penetration depth,
519 and thus inversely proportional to foot area and substrate strength (see Appendix).

520 *Granular surface model assumptions*

521 In our modeling of the foot-ground interaction on the granular surface using the penetration force
522 model, we made two assumptions. First, we assumed that the ground reaction forces were
523 insensitive to speed. This is true in the low speed regime (<0.5 m/s for our glass particles,
524 Maladen et al., 2009) where particle inertia is negligible and forces are dominated by particle
525 friction. Because friction is proportional to pressure, and pressure is proportional to depth (Hill et

526  al., 2005), granular forces in the low-speed regime are proportional to depth ($F_z=\alpha|z|A$),
527  analogous to the hydrostatic forces in fluids ($F_z=\rho g|z|A$, i.e. buoyant forces due to hydrostatic
528  pressure).

529  Second, we used the vertical stress per unit depth $\alpha$ determined from vertical penetration of a
530  horizontally oriented disc to estimate forces on the foot as it rotates subsurface. In this calculation,
531  the effective vertical stress per unit depth $\alpha\cos\theta_{foot}$ (see Appendix) depended on foot orientation
532  via a simple relation $\cos\theta_{foot}$ (because projected area $A=A_{foot}\cos\theta_{foot}$; see Appendix), and not on
533  direction of motion. However, our recent physics experiments (Li et al., in preparation) suggest
534  that stresses in granular media in the low speed regime depend on both orientation and direction
535  of motion in a more complicated manner, and that $\alpha\cos\theta_{foot}$ overestimates vertical stress per unit
536  depth for all foot orientations and directions of motion except when the foot is horizontal and
537  moving vertically downwards. Therefore, our model must be overestimating hydrostatic-like
538  forces, and there must be additional forces contributing to the lizard's ground reaction forces.

539  We propose that these additional forces are likely from hydrodynamic-like inertial forces
540  resulting from the local acceleration of the substrate (particles) by the foot. Analogous to
541  hydrodynamic forces in fluids (Vogel, 1996), for an intruder moving rapidly in granular media,
542  the particles initially at rest in front of the intruder are accelerated by, and thus exert reaction
543  forces on, the intruder. Hydrodynamic-like forces at ~1 m/s can play an important role both in
544  impact forces on free falling intruders (Katsuragi and Durian, 2007; Goldman and Umbanhowar,
545  2008) and in legged locomotion of small lightweight robots (Qian et al., 2012). We note that the
546  foot rotation hypothesis should hold regardless, because hydrodynamic-like forces are also
547  proportional to projected area (Katsuragi and Durian, 2007).

548  However, we know too little about the lizard's subsurface foot kinematics and the force laws in
549  the high-speed regime on an intruder being pushed in a complex path within granular media (not
550  simply a free-falling intruder) to more accurately calculate both hydrostatic-like and
551  hydrodynamic-like forces. Future x-ray high-speed imaging experiments (Maladen et al 2009;
552  Sharpe et al., 2012) will reveal how the lizard foot was moving subsurface. Further studies of
553  intrusion forces in granular media in both low-speed (Li et al., 2013) and high-speed regimes can
554  provide a more comprehensive understanding of ground reaction forces during legged locomotion
555  on granular surfaces and provide better estimates of foot penetration depth and energy loss.

556                              *Comparison to water-running in basilisk lizard*

The rapid impact of the foot on the surface at touchdown and hypothesized subsurface foot rotation appear kinematically similar to the slap and stroke phases of the basilisk lizards running on the surface of water (Glasheen and McMahon, 1996a; Hsieh, 2003). For the zebra-tailed lizard running on sand, both granular hydrostatic-like and hydrodynamic-like forces can contribute to vertical ground reaction force. This is also qualitatively similar to water-running basilisk lizard, which utilizes both hydrostatic forces resulting from the hydrostatic pressure between the water surface and the bottom of the air cavity created by the foot, and hydrodynamic forces resulting from the water being accelerated from rest by the rapidly moving foot (Glasheen and McMahon, 1996a, 1996b; Hsieh and Lauder, 2004).

However, the degree to which each species relies on these two categories of forces differs due to differences in the properties of the supporting media. For given foot size, depth, and speed, the hydrostatic(-like) forces in water are an order of magnitude smaller than the hydrostatic-like forces in granular media, whereas the hydrodynamic(-like) forces are similar between in water and in granular media (see Appendix). As a result, the basilisk lizard running on water must rely on hydrodynamic forces to a larger degree than the zebra-tailed lizard running on sand, considering that these two lizards have similar size ($\sim$ 0.1 m). An extreme example for this is that it is impossible for a basilisk lizard to stand on the surface of water, but a zebra-tailed lizard can stand on loose sand.

*Motor function of upper hind leg*

Despite the passive nature of the leg spring in the spring-mass model, animal limbs do not function purely passively as springs—the muscles within them must perform mechanical work. We have shown that on the solid surface, the lizard's hind foot saves about 40% of the mechanical work per step. The remaining 60% is lost either within the foot or to the ground, and must be compensated by mechanical work performed by muscles, which is $W_{muscle} = 0.11 \pm 0.10$. This work is likely provided by knee extension during the second half of stance (Fig. 6D, red curve) powered by the upper leg muscles.

On the granular surface, substantial energy is lost to the substrate. This is in accord with previous observations of higher mechanical energetic cost during locomotion on granular surfaces in human (Zamparo et al., 1992; Lejeune et al., 1998) and legged robots (Li et al., 2010a). Because the energy lost to the substrate equals the reduction in CoM mechanical energy during the first half of stance, even without energy loss within the limb, the upper hind leg muscles must perform mechanical work of $W_{muscle} = 0.31 \pm 0.10$ during the second half of stance, about three times that

589    on the solid surface for a given animal running at a given speed, as evidenced by the larger knee

590    extension on the granular surface (Fig. 6D, blue curve).

591    Our models of the foot-ground interaction on both surfaces assume purely passive foot mechanics,

592    and do not consider the role of active neurosensory control. However, animals can actively adjust

593    kinematics and muscle function to accommodate changes in surface conditions (Ferris et al., 1999;

594    Daley and Biewener, 2006). We observed that when confronted by a substrate which transitioned

595    from solid into granular (or *vice versa*), the lizard displayed partial adjustment of foot posture

596    during the first step on the new surface, followed by full adjustment during the second step.

597    Future studies using neuromechanics techniques, such as EMG (Biewener et al., 1998; Sponberg

598    and Full, 2008; Sharpe and Goldman, in review) and denervation and reinnervation (Chang et al.,

599    2009), can determine how neural control and sensory feedback mechanisms are used to control

600    limb function to accommodate changing substrates.

601                                                    *Conclusions*

602    During running on both solid and granular surfaces, the zebra-tailed lizard displayed spring-mass-

603    like center of mass kinematics with distinct hind foot, hind leg, and trunk kinematics. The lizard's

604    large, elongate hind foot served multiple functions during locomotion. On the solid surface, the

605    hind foot functioned as an energy-saving spring and reduced about 40% of the mechanical work

606    needed each footstep. On the granular surface, the hind foot paddled through fluidized grains to

607    generate force, and substantial energy was lost during irreversible deformation of the granular

608    substrate. The energy lost within the foot and to the substrate must be compensated for by

609    mechanical work done by the upper hind leg muscles.

610    The multifunctional hind foot may passively (and possibly actively) adjust to the substrate during

611    locomotion in natural terrain, and provide this desert generalist with energetic advantages and

612    simplify its neurosensory control tasks (Full and Koditschek, 1999). Current robotic devices often

613    suffer performance loss and high cost of transport on flowing substrates like granular material

614    (Kumagai 2004; Li et al., 2009; Li et al., 2010b; Li et al., 2010a). Insights from studies like ours

615    can provide inspiration for next-generation multi-terrain robots (Pfeifer et al., 2007). Finally, our

616    study also highlights the need for comprehensive force models for granular media (Li et al., in

617    preparation) and for flowing terrestrial environments in general.

618

619                                                 **Appendix**

620                                  *Small relative oscillation in forward speed*

621   Running at 1.1 m/s, the lizard's Froude number in the sagittal plane was $Fr = v_{x,\text{CoM}}^2/gL_0 = 3$

622   (where $L_0 \approx 4$ cm is the leg length at touchdown), above the typical value of 2.5 where most

623   animals transition from trotting to galloping (Alexander, 2003). This implied that the kinetic

624   energy ($\frac{1}{2} mv_{\text{CoM}}^2 \approx \frac{1}{2} mv_{x,\text{CoM}}^2$) of the CoM was 3 times larger than its gravitational potential

625   energy ($mgz_{\text{CoM}}$). Because both the forward speed oscillation $\Delta v_{x,\text{CoM}}$ and vertical speed oscillation

626   $\Delta v_{z,\text{CoM}}$ were determined by the total ground reaction force and the attack angle of the leg spring

627   ($\beta = \sin^{-1}(v_{x,\text{CoM}}DT/2L_0) = 0.9$ rad), they must be of the same order of magnitude (Blickhan, 1989),

628   i.e., $\Delta v_{x,\text{CoM}} \sim \Delta v_{z,\text{CoM}}$. From the observed CoM kinematics, $\Delta v_{z,\text{CoM}} < (gL_0)^{1/2}$. Therefore, $\Delta v_{x,\text{CoM}} \sim$

629   $\Delta v_{z,\text{CoM}} < (gL_0)^{1/2} << v_{x,\text{CoM}}$, and $\Delta v_{x,\text{CoM}}/v_{x,\text{CoM}} << 1$.

630                                  *Hind foot curvature on solid surface*

631   Three-dimensional kinematics showed that the hind limb (from the hip to the digit tip of the

632   fourth toe) remained nearly within a plane during the entire stride (out-of-plane component is 5%

633   averaged over the entire stride). During stance, the orientation of the foot plane remained nearly

634   unchanged, with a foot sprawl angle of $53 \pm 4°$ relative to the sagittal plane in the posterior view.

635   Hind foot curvature $\kappa$ could then be obtained by fitting a circle to the hind foot (from the ankle to

636   the digit tip) within the foot plane and determining the radius of curvature $\rho$ of the fit circle (see

637   diagram in Fig. 9A), i.e., $\kappa = \pm 1/\rho$, where + sign indicates foot hyperextension, − sign indicates

638   foot flexion, and $\kappa = 0$ indicates a straight foot.

639                                  *Tendon spring deformation*

640   From the two-dimensional strut-spring model of the hind limb, by geometry, the tendon spring

641   deformation $\Delta l$ was related to the observed changes of joint angles and the foot joint radii as: $\Delta l =$

642   $\Sigma_i r_i \Delta\theta_i$, where $i = $ K, A, MP, PP were the four joints in the model, $\Delta\theta_i$ the observed changes of

643   joint angles, and $r_i$ the joint radii ($r_K = r_A = 1.25$ mm, $r_{MP} = 0.75$ mm, $r_{PP} = 0.50$ mm). We

644   observed that the relaxed hind foot of a live animal was nearly straight (Fig. 1A), which was

645   similar to the foot shape at touchdown during running (Fig. 3A,E). Thus we defined the relaxed

646   length of the tendon spring as the length when the foot was straight, i.e., $\Delta l = 0$ at touchdown.

647   Calculated maximal tendon spring deformation $\Delta l_{\max} = 0.78$ mm corresponded to a 3% strain. We

648   did not consider tendon spring deformation in the swing phase (dotted curve in Fig. 6F) because

649   the assumption of isometric contraction of lower leg muscles was only valid for the stance phase.

650                                 *Tendon spring stiffness*

651   The stiffness of the tendon spring was defined as the maximal tension divided by the maximal

652   deformation of the tendon spring, i.e., $k = T_{max}/\Delta l_{max}$. From the observed CoM kinematics, the

653   total ground reaction force at mid-stance was $F_{max} = 0.3$ N within the coronal plane and pointed

654   from the digit tip to the hip. At mid-stance, because the foot was neither dorsiflexing nor

655   plantarflexing, torque was balanced at the ankle, i.e., $T_{max}r_A = F_{max}\Delta x_{AT}$, where $\Delta x_{AT} = 1.4$ cm was

656   the horizontal distance between the ankle and the digit tip at mid-stance, and $r_A = 1.25$ mm. Thus

657   $T_{max} = 3.4$ N and $k = 4.4 \times 10^3$ N/m. The maximal stress in the foot tendons during stance was

658   $\sigma_{max} = T_{max}/\pi r_{PP}^2 = 4.3$ MPa.

659   The torsional stiffness of the ankle observed in anesthetized lizards from the modified work loop

660   experiments (~ $1 \times 10^{-3}$ Nm/rad) was an order of magnitude smaller than estimated from running

661   kinematics ($12 \times 10^{-3}$ Nm/rad). This is however not contradictory but expected because during

662   stance the lizard's lower leg muscles must be activated, and the resulting higher tension from

663   muscle contraction increases limb stiffness (Weiss et al., 1988).

664                         *Foot elongation increases energy savings on solid surface*

665   The stiffness of a piece of elastic material like a tendon is $k = E_0A_0/l_0$, where $E_0$ is the Young's

666   modulus, $A_0$ the cross sectional area, and $l_0$ the rest length of the material. Most animal tendons

667   are primarily made of collagen (Kirkendall and Garrett, 1997) and are of similar Young's

668   modulus (i.e., $E_0$ is nearly constant). Thus, the stiffness of the tendon spring scales as $k \propto A_0/l_0 \propto$

669   $r_0^2/l_0$, i.e., an elongate tendon (smaller $r_0$ and larger $l_0$) is less stiff and stretches more easily than

670   a short, thick tendon. Because elastic energy storage decreases with tendon stiffness ($E_{storage} = ½$

671   $k\Delta l_{max}^2 = ½ T_{max}^2/k \propto 1/k$ for a given $T_{max}$), an elongate tendon can store (and return) more energy.

672   An elongate foot also reduces the moment arm of tendon tension (small $r_A$) but increases the

673   moment arm of the ground reaction force (large $\Delta x_{AT}$) about the ankle, therefore reducing the

674   mechanical advantage (Biewener et al., 2004), so it increases tension in the foot for a given

675   ground reaction force (because $T_{max} = F_{max}\Delta x_{AT}/r_A$) and amplifies tendon stretch for enhanced

676   energy storage and return.

677                         *Vertical ground reaction force on granular surface*

678   We assumed that the hind foot was rotating at a constant angular velocity ω about the moving

679   ankle during stance, i.e., $\theta_{foot} = \omega t$ within $0 \leq t \leq DT$ and $0 \leq \theta_{foot} \leq \pi/2$, then $\omega = \pi/2DT = 10\pi/9T$

680    = 35 rad/s. From the measured vertical speed of the ankle and this assumed foot rotation, the

681    vertical speed of most (75%) of the foot was always below 0.5 m/s during most (75%) of stance.

682    Given foot rotation $\theta_{foot} = \omega t$, the foot area projected in the horizontal plane decreased with time

683    as $A = A_{foot}\cos\omega t$, where $A_{foot} = 1$ cm$^2$ is the hind foot area; the foot depth (measured at the center

684    of the foot) increased with time as $|z| = |z|_{max}\sin\omega t$. The vertical ground reaction force on the foot

685    was then sinusoidal: $F_z = F_{z,max}\sin 2\omega t$, which was sinusoidal, where $F_{z,max} =$

686    $\alpha A_{foot}|z|_{max}\sin\pi/4\cos\pi/4 = \frac{1}{2}\alpha A_{foot}|z|_{max}$. For steady state locomotion on a level surface, the $F_z$

687    generated by one foot averaged over a cycle must equal half the body weight, i.e.,

688    $\int_0^T F_{z,max}\sin 2\omega t\, dt = \frac{1}{2}mg$. Therefore, $F_{z,max} = 5\pi mg/9$ and $F_z = 5\pi mg/9 \sin 10\pi t/9T$.

689                                *Energy loss to granular substrate*

690    By integration of vertical ground reaction force over vertical displacement of the foot, the energy

691    loss to the granular substrate was $E_{substrate} = \int_0^{|z|_{max}} F_z\, d|z| = \int_0^T F_z \frac{d|z|}{dt}\, dt$, where $|z|_{max} = 1.0$ cm

692    from $F_{z,max} = \frac{1}{2}\alpha A_{foot}|z|_{max}$. The hypothesized foot rotation in the sagittal plane did not take into

693    account the sprawl of the foot during stance, which could induce additional energy loss by lateral

694    displacement of the granular substrate. However, a sprawled foot posture did not affect the

695    condition of vertical force balance and thus did not change our estimate of energy dissipation in

696    the sagittal plane. Therefore this estimate provides a lower bound.

697                *Large foot area reduces energy loss on granular surface*

698    For a given animal (constant weight $mg$), $F_{z,max} = \frac{1}{2}\alpha A_{foot}|z|_{max} = 5\pi mg/9$ is constant, thus $E_{substrate}$

699    $= |z|_{max}\int_0^T F_z\ \omega\cos\omega t\, dt \propto |z|_{max} \propto 1/(\alpha A_{foot})$. This implies that the energy loss to the granular

700    substrate increases with foot penetration depth. On a given granular surface (fixed $\alpha$), a larger

701    foot (larger $A_{foot}$) sinks less than a smaller foot, and thus loses less energy to the substrate. For a

702    given foot size (fixed $A_{foot}$), a foot sinks less on a stronger granular substrate (larger $\alpha$) than on a

703    weaker substrate, and thus loses less energy to the substrate.

704                *Comparison of forces in granular media and in water*

705    For water, hydrostatic force is $F_z = \rho g|z|A$. Comparing this with $F_z = \alpha|z|A$ for granular media, $\rho g$

706    is the equivalent of $\alpha$. For water, $\rho g = 1.0 \times 10^4$ N/m$^3$; for loosely packed glass particles, $\alpha = 3.5$

707    $\times 10^5$ N/m$^3$. Therefore, the hydrostatic forces in water are an order of magnitude smaller than the

708    hydrostatic-like forces in granular media for given foot size and depth.

709 Hydrodynamic(-like) forces should be proportional to the density of the surrounding media
710 because they are due to the media being accelerated. For water, $\rho = 1.0 \times 10^3$ N/m$^3$; for loosely
711 packed glass particles the effective density is $2.5 \times 10^3$ N/m$^3 \times 0.58$ volume fraction $= 1.45 \times 10^3$
712 N/m$^3$. Therefore, the hydrodynamic forces in water and hydrodynamic-like forces in granular
713 media are on the same order of magnitude for given foot size and foot speed.

714

715 **Acknowledgements**

716 We gratefully thank Sarah Sharpe, Yang Ding, Nick Gravish, Ryan Maladen, Paul Umbanhowar,
717 Kyle Mara, Young-Hui Chang, Andy Biewener, Tom Roberts, Craig McGowan, and two
718 anonymous reviewers for helpful discussions and/or comments on the manuscript; Loretta Lau for
719 help with kinematics data tracking; Sarah Sharpe for help with animal protocol and
720 anesthetization; Mateo Garcia, Nick Gravish, and Andrei Savu for help with force plate setup;
721 Ryan Maladen and The Sweeney Granite Mountains Desert Research Center for help with animal
722 collection; and the staff of The Physiological Research Laboratory animal facility of The Georgia
723 Institute of Technology for animal housing and care. This work was funded by The Burroughs
724 Wellcome Fund (D.I.G. and C.L.), The Army Research Laboratory Micro Autonomous Systems
725 and Technology Collaborative Technology Alliance (D.I.G. and C.L.), The Army Research
726 Office Biological Locomotion Principles and Rheological Interaction Physics (D.I.G. and C.L.)
727 and The University of Florida and Temple University start-up funds (S.T.H.).

728

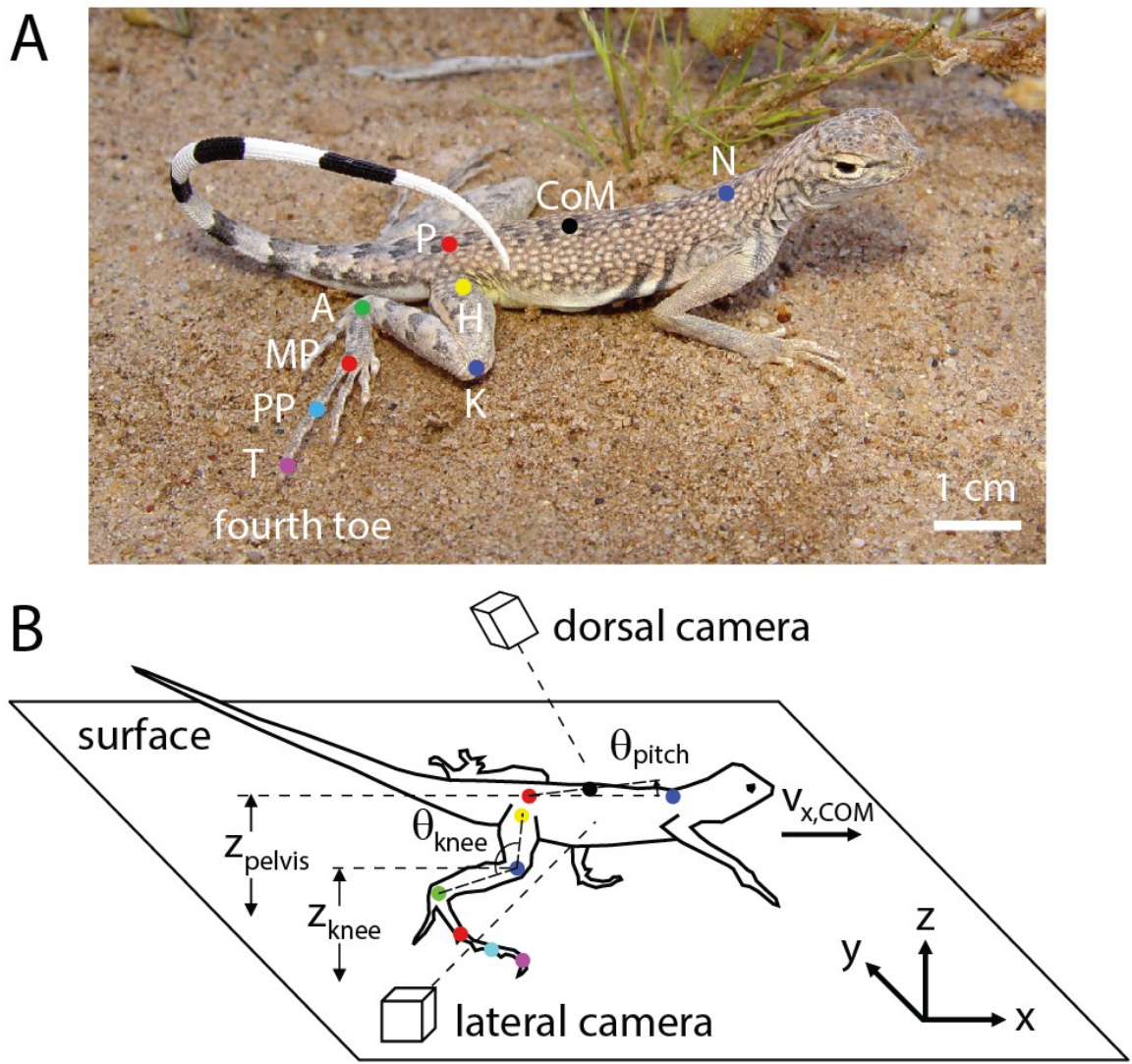

Fig. 1. Model organism and three-dimensional kinematics experiments. (A) A zebra-tailed lizard resting on sand in the wild (photo: Thomas C. Brennan). (B) Experimental setup for three-dimensional kinematics capture, with definitions of pelvis height ($z_{pelvis}$), knee height ($z_{knee}$), trunk pitch angle ($\theta_{pitch}$), and knee angle ($\theta_{knee}$). Colored dots in (A,B) are digitized points on the midline of the trunk, hind leg, and elongate hind foot.

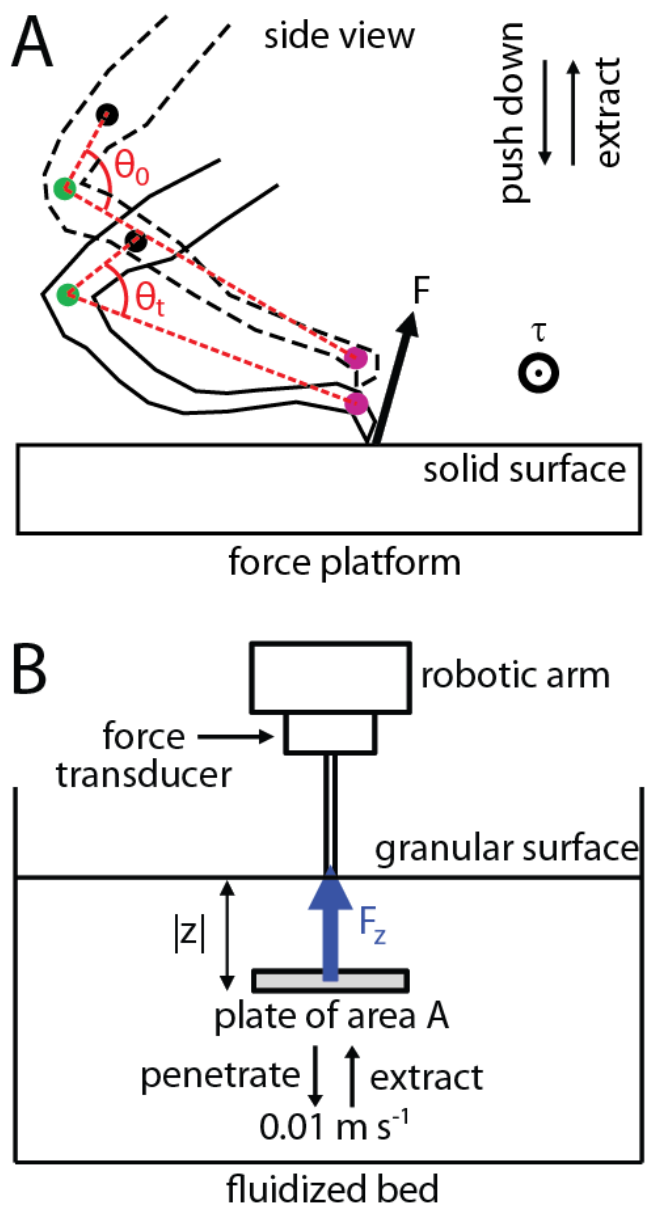

Fig. 2. Experiments to measure hind limb resilience and granular penetration force. (A) Experimental setup for hind limb resilience measurements. Dashed foot tracing shows the relaxed, straight foot right before touchdown. Solid foot tracing shows the hyperextended foot during ground contact. $F$, ground reaction force; $\theta_0$, angle between the ankle and the digit tip in the relaxed, straight foot; $\theta_t$, angle between the ankle and the digit tip in the hyperextended foot; $\tau$, torque about the ankle. (B) Experimental setup for granular penetration force measurements.

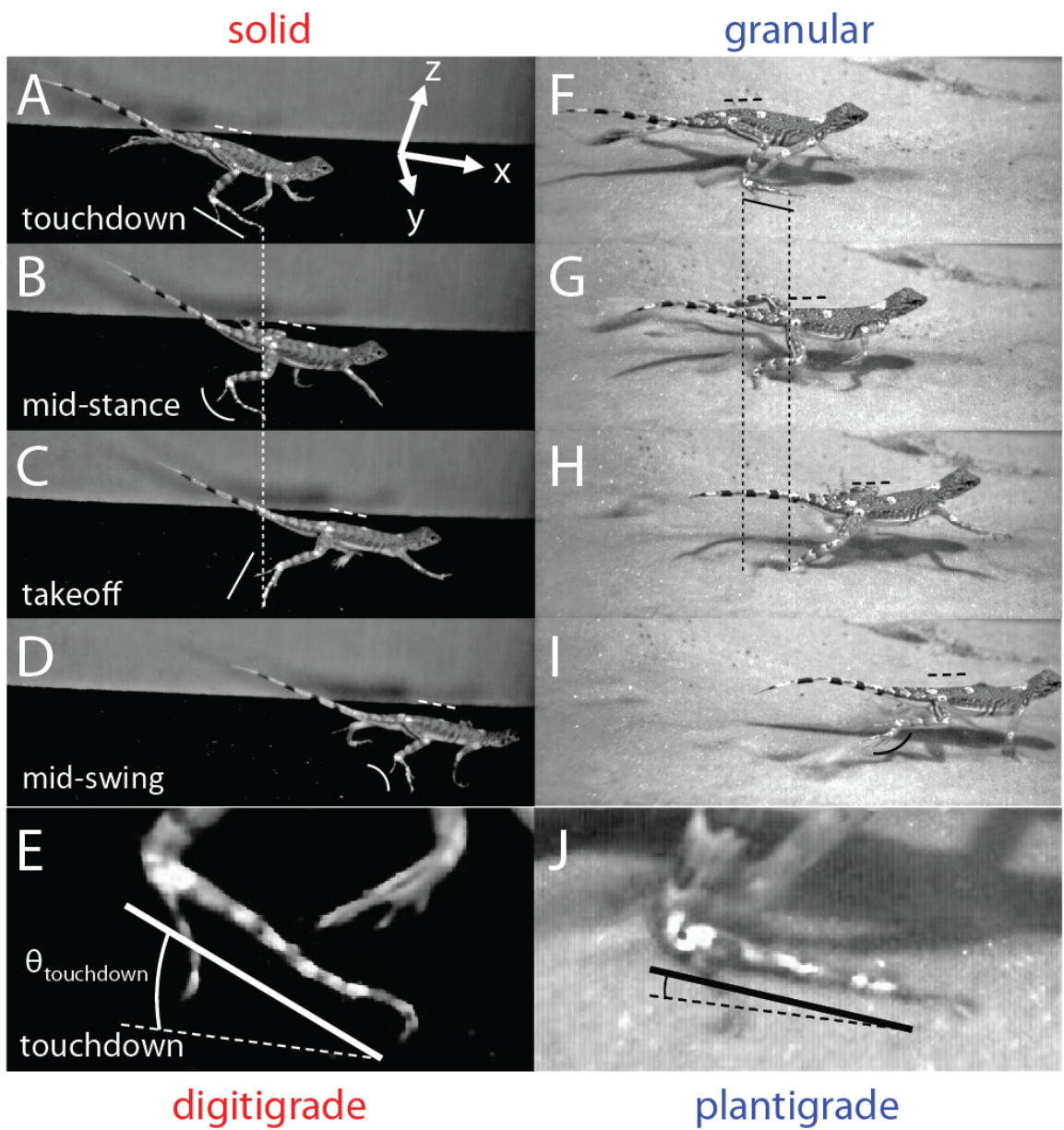

922

Fig. 3. Lateral views of representative runs on the solid (A–D) and the granular (F–I) surface (see Movies 1, 2 in supplementary material). (E,J) Closer views of foot posture at touchdown showing definition of touchdown foot angle $\theta_{touchdown}$. Solid lines and curves along the foot indicate hind foot posture and shape. Note that the lateral camera was oriented at an angle to the *x, y, z* axes such that forward (+*x*) direction appeared to point slightly downwards.

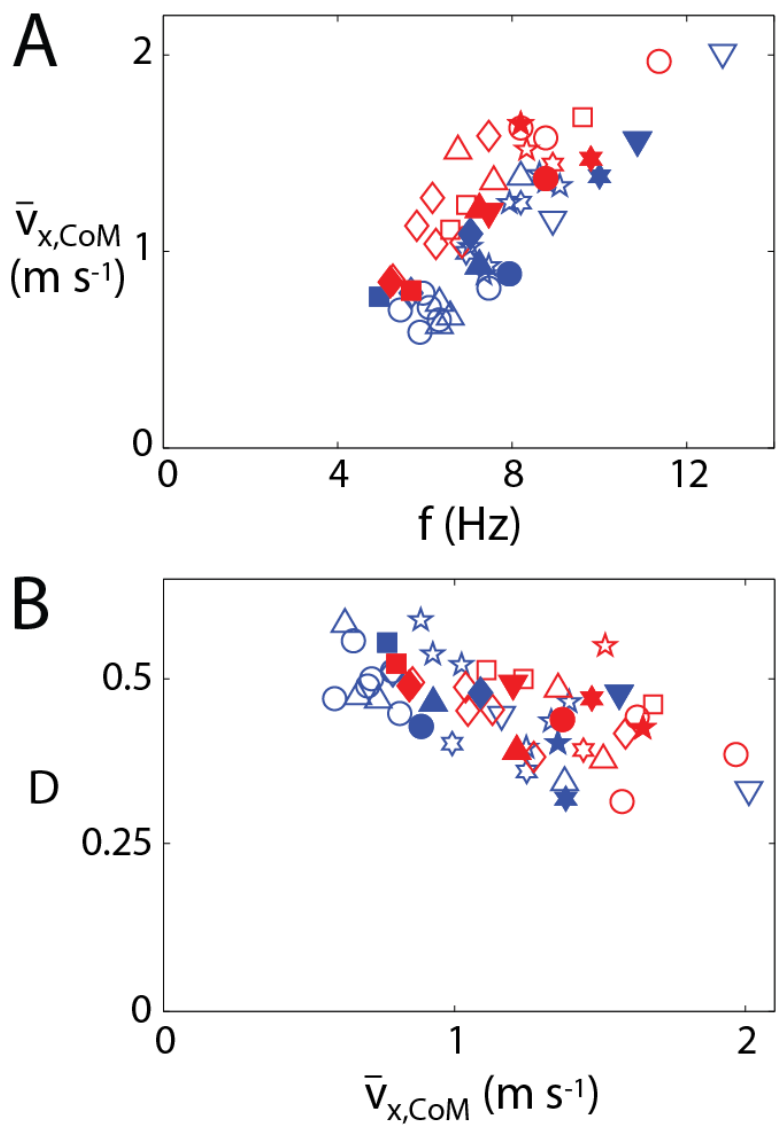

928

929  Fig. 4. Performance and gait on the solid (red) and the granular (blue) surfaces. (A) Average
930  forward speed vs. stride frequency. (B) Duty factor vs. average forward speed. Different symbols
931  represent different individuals. Filled symbols are from the seven representative runs for each of
932  the seven individuals tested on both substrates. Empty symbols are from runs that were not
933  included in the representative data set.

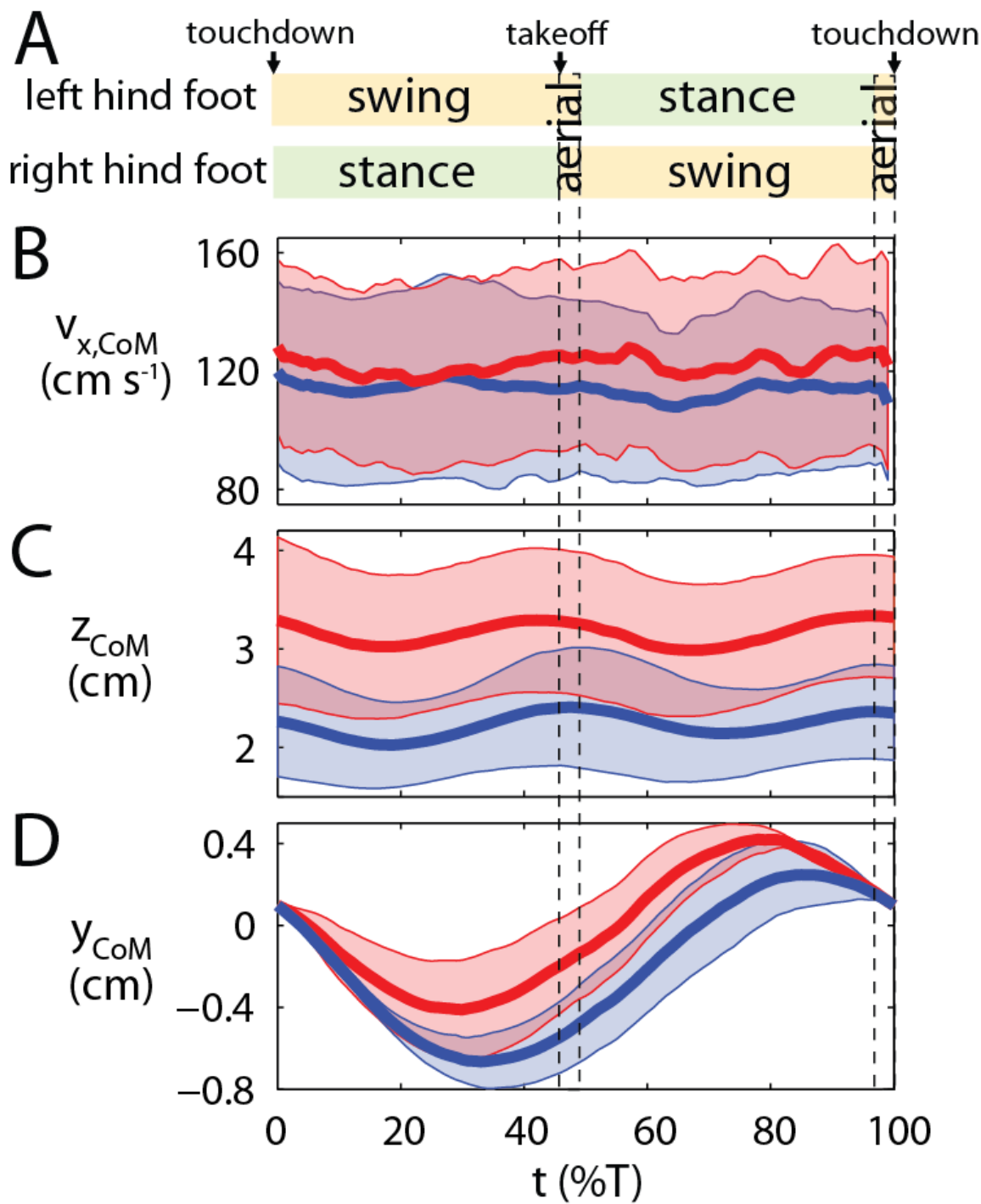

935 Fig. 5. Center of mass (CoM) kinematics (mean ± s.d.) vs. time during a stride on the solid (red)

936 and the granular (blue) surfaces. (A) Footstep pattern. (B) CoM forward speed. (C) CoM vertical

937 position. (D) CoM lateral position. See Fig. 1 for definitions of kinematic variables.

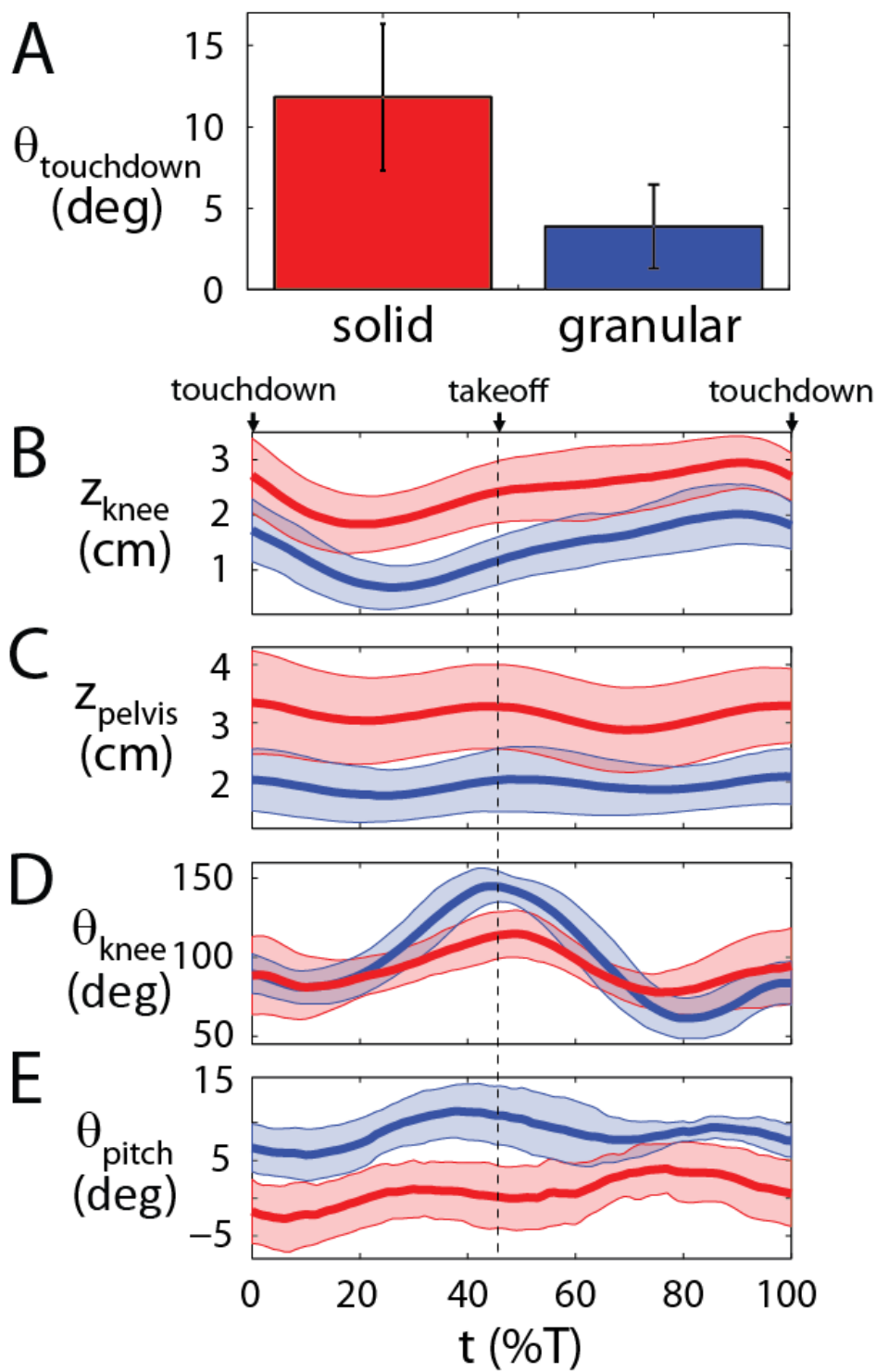

938

939  Fig. 6. Hind foot, hind leg, and trunk kinematics (mean ± s.d.) vs. time during a stride on the solid
940  (red) and the granular (blue) surfaces. (A) Touchdown foot angle. (B) Knee height. (C) Pelvis
941  height. (D) Knee angle. (E) Trunk pitch angle. See Fig. 1 and Fig. 3E,J for definitions of
942  kinematic variables.

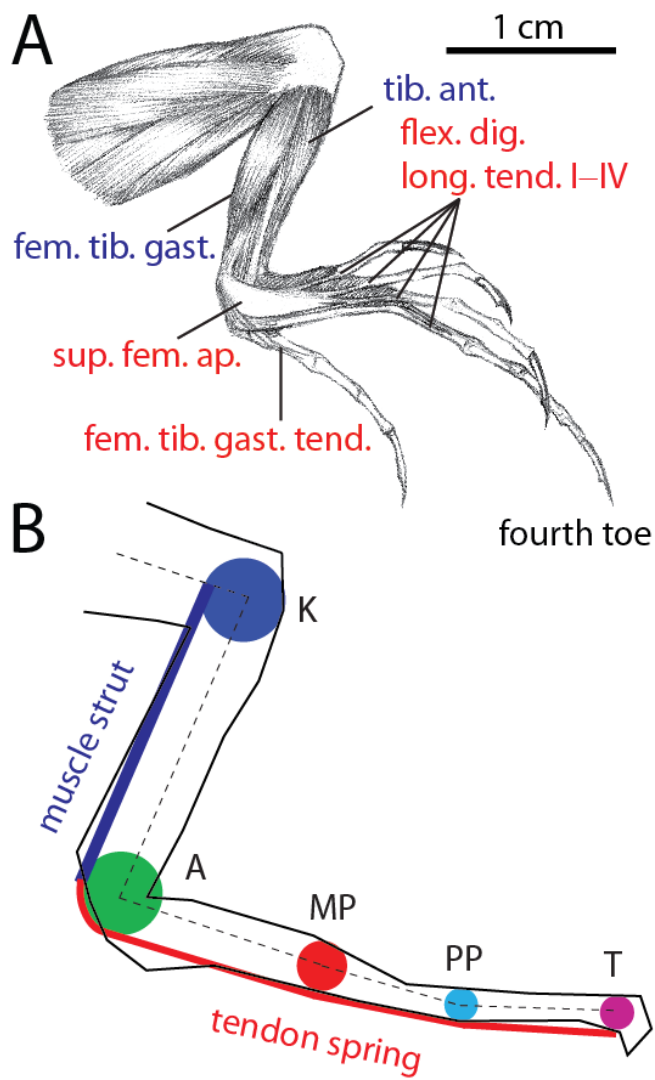

943

Fig. 7. Anatomy and a strut-spring model of the hind limb. (A) Ventral anatomy of a dissected hind limb. Lower hind leg muscles are marked in blue; foot tendons are marked in red. (B) A two-dimensional model of the hind limb. The muscle strut models isometrically contracting lower leg muscles; the tendon spring models foot tendons. The radii of colored circles correspond to measured joint radii.

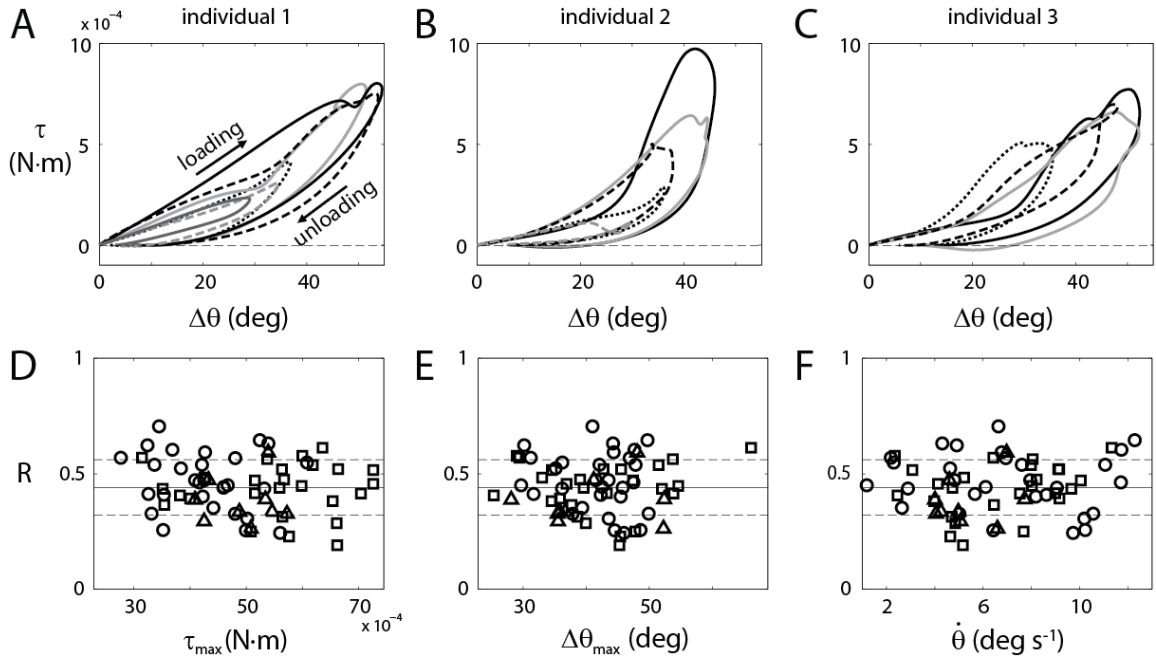

Fig. 8. Hind limb resilience. (A–C) Representative passive work loops of the hind foot (measured at the digit tip) from each of the three anesthetized lizards tested. Different curves are from different trials. The area within a work loop is the energy lost within the foot. See Fig. 2A for schematic of experimental setup. (D–F) Hind limb resilience vs. maximal torque, maximal angular displacement, and average loading rate. Different symbols are from different individuals. Solid and dashed lines in (D–F) denote mean ± s.d.

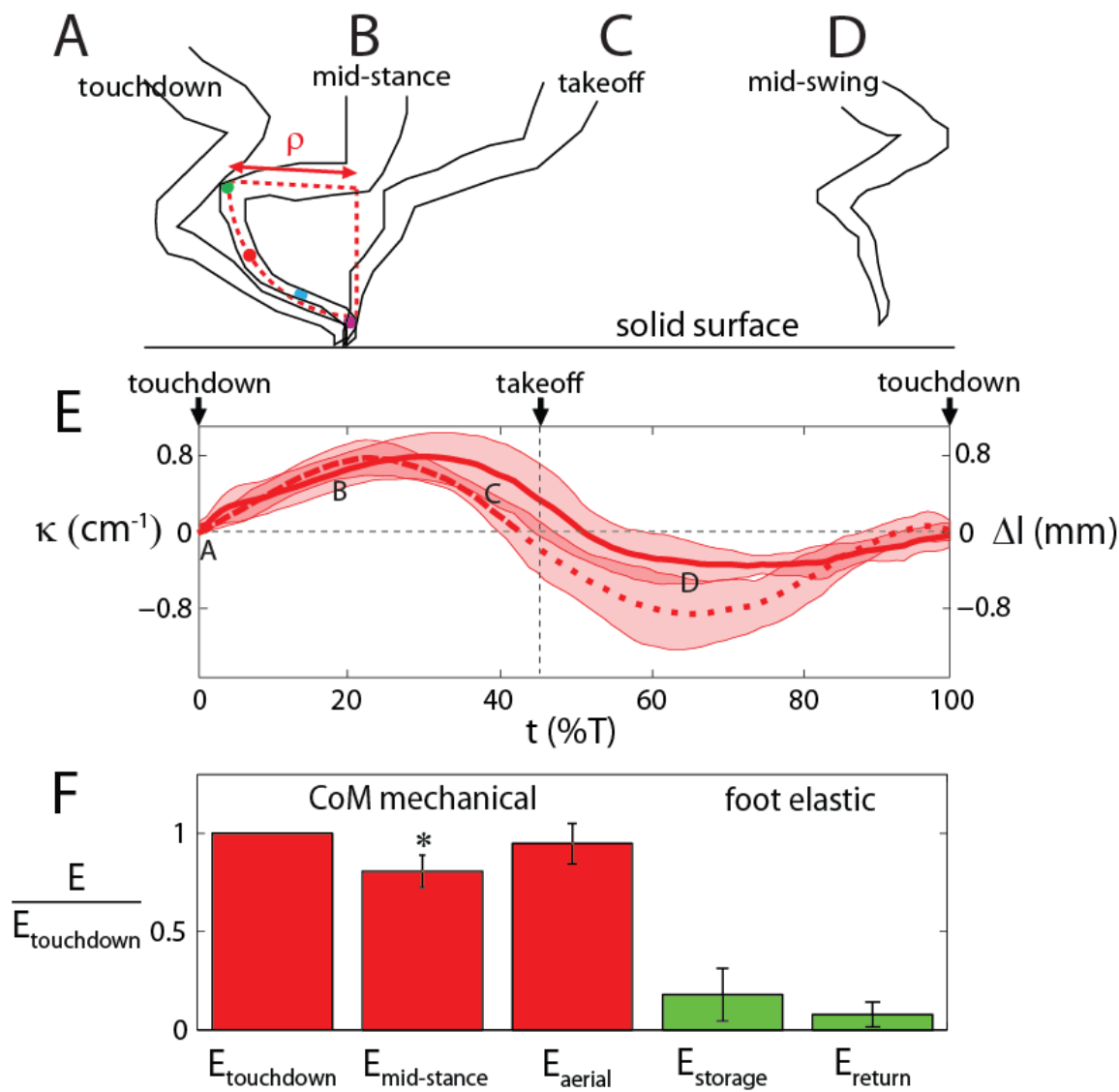

Fig. 9. Foot-ground interaction on the solid surface. (A–D) The hind foot shape from the lateral view of a representative run on the solid surface. (A–D) correspond with (A–D) in Fig. 3. The hind foot shape in the dorsal view is similar because the sprawl angle of the foot plane is nearly constant during stance. The diagram in (B) defines the radius of curvature $\rho$ of the foot (see Appendix). (E) Foot curvature (solid) and tendon spring deformation (dashed) (mean ± s.d.) vs. time during a stride on the solid surface. Tendon spring deformation is not meaningful during swing (dotted) when the muscle strut assumption does not hold. (F) Mechanical energies of the CoM and elastic energies of the foot (mean ± s.d.) on the solid surface. All energies are normalized to the mechanical energy of the CoM at touchdown ($E_{touchdown}$) for each run. * indicates that $E_{mid-stance}$ is significantly different from $E_{touchdown}$ and $E_{aerial}$ ($P < 0.05$, ANOVA, Tukey HSD).

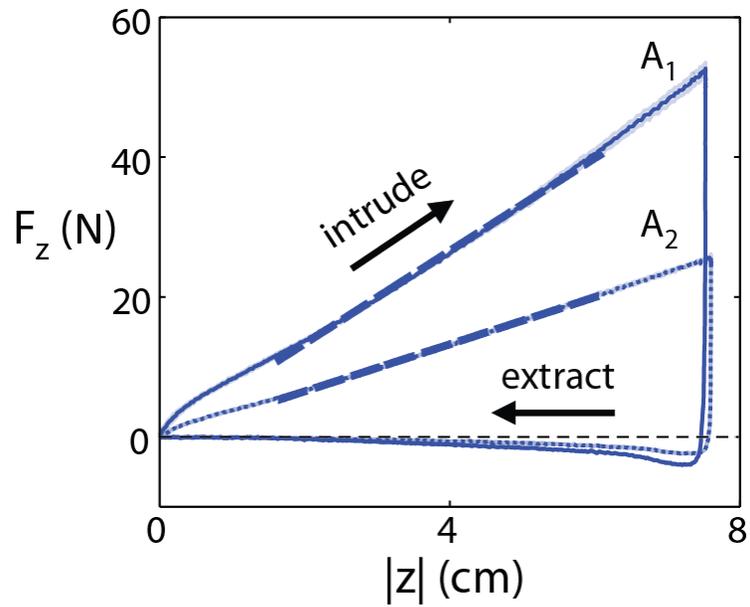

970

Fig. 10. Granular penetration force (mean ± s.d.) vs. depth on two plates of different areas: $A_1$ =
7.6 × 2.5 cm$^2$ and $A_2$ = 3.8 × 2.5 cm$^2$. See Fig. 2B for schematic of experimental setup. Dashed
lines are linear fits to the data over steady state during penetration using Eqn. (1).

974

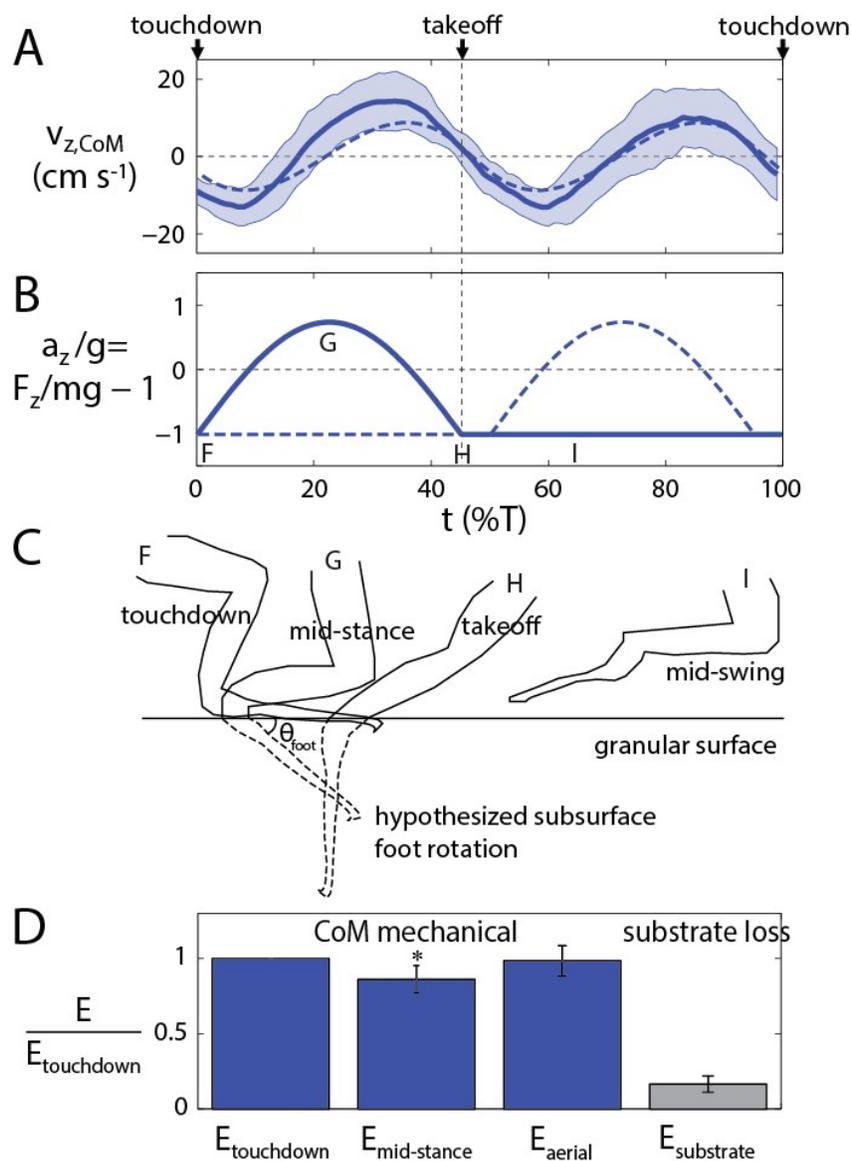

Fig. 11. Foot-ground interaction on the granular surface. (A) CoM vertical speed (mean ± s.d.) vs. time during a stride. Solid curve is from experiment. Dashed curve is calculated from the vertical acceleration from the model. (B) Vertical acceleration vs. time during a stride calculated from the total vertical ground reaction force $F_z$ on both feet and the animal weight $mg$. Solid and dashed curves are the $F_z$ on the two alternating hind feet. (C) Hypothesized subsurface foot rotation in the sagittal plane. (F–I) correspond with (F–I) in Fig. 3. $\theta_{foot}$, foot angle in the vertical plane. (D) Mechanical energy of the CoM and the energy loss to the substrate (mean ± s.d.) during running on the granular surface. All energies are normalized to the mechanical energy of the CoM at touchdown ($E_{touchdown}$) for each run. * indicates that $E_{mid\text{-}stance}$ is significantly different from $E_{touchdown}$ and $E_{aerial}$ ($P < 0.05$, ANOVA, Tukey HSD).

987     Table 1. Morphological measurements (mean ± s.d.) of the seven individuals tested in the 3-D

988     kinematics experiments.

| | |
|---|---|
| SVL (cm) | 7.2 ± 0.6 |
| Mass *m* (g) | 11.0 ± 2.7 |
| Trunk length (cm) | 4.4 ± 0.4 |
| Pelvic width (cm) | 1.4 ± 0.1 |
| Hind limb length (cm) | 6.4 ± 0.1 |
| Hind foot length (cm) | 2.7 ± 0.1 |
| Femur length (cm) | 1.6 ± 0.2 |
| Tibia length (cm) | 2.1 ± 0.2 |
| Tarsals and metatarsals length (cm) | 1.0 ± 0.1 |
| Fourth toe length (cm) | 1.7 ± 0.1 |

989

990 Table 2. Gait and kinematic variables (mean ± s.d.) and statistics using an ANCOVA. *P* values
991 reported are for substrate effect.

| Variable | Solid | Granular | *F* | *P* |
|---|---|---|---|---|
| †Average forward speed $\bar{v}_{x,\mathrm{CoM}}$ (m/s) | 1.2 ± 0.3 | 1.1 ± 0.3 | 0.4784 | 0.5023 |
| Stride frequency $f$ (Hz) | 7.5 ± 1.6 | 8.1 ± 2.0 | 9.9101 | **0.0319** |
| Duty factor $D$ | 0.46 ± 0.05 | 0.45 ± 0.07 | 0.5032 | 0.5480 |
| Stride length $\lambda$ (m) | 0.16 ± 0.02 | 0.14 ± 0.02 | 8.9112 | **0.0409** |
| Average CoM height $\bar{z}_{\mathrm{CoM}}$ (cm) | 3.2 ± 0.7 | 2.2 ± 0.5 | 5.4690 | **0.0203** |
| Magnitude of CoM vertical oscillations $\Delta z_{\mathrm{CoM}}$ (cm) | 0.3 ± 0.2 | 0.4 ± 0.3 | 3.7031 | 0.4697 |
| Lowest CoM height $z_{\mathrm{CoM}}$ (cm) | 3.0 ± 0.7 | 2.0 ± 0.4 | 7.7544 | **0.0115** |
| Time of lowest CoM height ($T$) | 0.18 ± 0.04 | 0.19 ± 0.04 | 0.9696 | 0.6366 |
| Highest CoM height $z_{\mathrm{CoM}}$ (cm) | 3.3 ± 0.7 | 2.4 ± 0.6 | 3.6126 | **0.0447** |
| Time of highest CoM height ($T$) | 0.44 ± 0.04 | 0.48 ± 0.01 | 3.0642 | **0.0325** |
| Magnitude of CoM lateral oscillations $\Delta y_{\mathrm{CoM}}$ (cm) | 0.86 ± 0.19 | 0.94 ± 0.23 | 0.2350 | 0.5263 |
| Average pelvis height pelvis $\bar{z}_{\mathrm{pelvis}}$ (cm) | 3.1 ± 0.7 | 1.9 ± 0.5 | 8.8912 | **0.0046** |
| Average trunk pitch angle $\bar{\theta}_{\mathrm{pitch}}$ (deg) | 1 ± 3 | 9 ± 2 | 19.5282 | **0.0002** |
| Touchdown knee height $z_{\mathrm{knee}}$ (cm) | 2.7 ± 0.7 | 1.7 ± 0.6 | 6.7157 | **0.0171** |
| Lowest knee height $z_{\mathrm{knee}}$ (cm) | 1.8 ± 0.5 | 0.7 ± 0.4 | 15.4261 | **0.0006** |
| Knee vertical displacement during stance $\Delta z_{\mathrm{knee}}$ (cm) | 0.9 ± 0.2 | 1.1 ± 0.4 | 0.7128 | 0.3056 |
| Touchdown knee angle $\theta_{\mathrm{knee}}$ (deg) | 88 ± 25 | 90 ± 13 | 1.2344 | 0.6713 |
| Lowest knee angle $\theta_{\mathrm{knee}}$ (deg) | 79 ± 17 | 79 ± 10 | 1.3175 | 0.7549 |
| Highest knee angle $\theta_{\mathrm{knee}}$ (deg) | 116 ± 15 | 150 ± 8 | 17.568 | **0.0001** |
| Knee joint extension during stance $\Delta\theta_{\mathrm{knee}}$ (deg) | 37 ± 13 | 71 ± 4 | 18.0994 | **0.0001** |
| ‡Average leg sprawl angle during stance $\theta_{\mathrm{sprawl}}$ (deg) | 40 ± 1 | 38 ± 5 | N/A | N/A |
| Touchdown foot angle $\theta_{\mathrm{touchdown}}$ (deg) | 12 ± 4 | 4 ± 3 | 7.6973 | **0.0032** |

992 All significant differences ($P < 0.05$) are in bold. Degree of freedom is (2,11) for all variables.

993 † An ANOVA was used to test the effect of substrate on running speed.

994 ‡ A direct comparison was not possible for $\theta_{\mathrm{sprawl}}$ between substrates because $\theta_{\mathrm{sprawl}}$ was measured
995 differently: on the solid surface, leg orientation was measured from the hip to the digit tip; on the
996 granular surface, leg orientation was measured from the hip to the ankle.

997

998    Table 3. Normalized energetic variables (mean ± s.d.). All energies were normalized to $E_{\text{touchdown}}$

999    for each run and averaged over 7 representative runs on each substrate.

| Variable | Solid | Granular |
|---|---|---|
| Mechanical energy at touchdown $E_{\text{touchdown}}$ | $1.00 \pm 0.00$ | $1.00 \pm 0.00$ |
| Mechanical energy at mid-stance $E_{\text{mid-stance}}$ | $0.81 \pm 0.08*$ | $0.86 \pm 0.09*$ |
| Mechanical energy during aerial phase $E_{\text{aerial}}$ | $0.95 \pm 0.10$ | $0.99 \pm 0.10$ |
| Mechanical energy reduction $\Delta E_{\text{mech}}$ | $0.19 \pm 0.08$ | $0.14 \pm 0.09$ |
| Elastic energy storage at mid-stance $E_{\text{storage}}$ | $0.18 \pm 0.13$ | N/A |
| Elastic energy return $E_{\text{return}}$ | $0.08 \pm 0.06$ | N/A |
| Energy loss to substrate $E_{\text{substrate}}$ | N/A | $0.17 \pm 0.05$ |
| Muscle mechanical work $W_{\text{muscle}}$ | $0.11 \pm 0.10$ | $0.31 \pm 0.10$ |

1000    * indicates significant difference ($P < 0.05$) in the mechanical energy of the CoM at mid-stance

1001    from that at touchdown and during aerial phase.